\documentclass{JHEP3}
\usepackage{graphicx}
\usepackage{epstopdf}

\newcommand{\be}{\begin{equation}}
\newcommand{\ee}{\end{equation}}
\newcommand{\br}{\begin{eqnarray}}
\newcommand{\er}{\end{eqnarray}}
\newcommand{\ba}{\begin{array}}
\newcommand{\ea}{\end{array}}
\newcommand{\bi}{\begin{itemize}}
\newcommand{\ei}{\end{itemize}}
\newcommand{\bn}{\begin{enumerate}}
\newcommand{\en}{\end{enumerate}}
\newcommand{\bc}{\begin{center}}
\newcommand{\ec}{\end{center}}

\def\epem{\ifmmode{e^+ e^-} \else{$e^+ e^-$} \fi}

\newcommand{\Dir}{\kern -6.4pt\Big{/}}
\newcommand{\Dirin}{\kern -10.4pt\Big{/}\kern 4.4pt}
\newcommand{\DDir}{\kern -8.0pt\Big{/}}
\newcommand{\DGir}{\kern -6.0pt\Big{/}}

\def\jhep #1 #2 #3 {{JHEP} {\bf#1} (#2) #3}
\def\plb #1 #2 #3 {{Phys.~Lett.} {\bf B#1} (#2) #3}
\def\npb #1 #2 #3 {{Nucl.~Phys.} {\bf B#1} (#2) #3}
\def\epjc #1 #2 #3 {{Eur.~Phys.~J.} {\bf C#1} (#2) #3}
\def\zpc #1 #2 #3 {{Z.~Phys.} {\bf C#1} (#2) #3}
\def\jpg #1 #2 #3 {{J.~Phys.} {\bf G#1} (#2) #3}
\def\prd #1 #2 #3 {{Phys.~Rev.} {\bf D#1} (#2) #3}
\def\prep #1 #2 #3 {{Phys.~Rep.} {\bf#1} (#2) #3}
\def\prl #1 #2 #3 {{Phys.~Rev.~Lett.} {\bf#1} (#2) #3}
\def\mpl #1 #2 #3 {{Mod.~Phys.~Lett.} {\bf#1} (#2) #3}
\def\rmp #1 #2 #3 {{Rev. Mod. Phys.} {\bf#1} (#2) #3}
\def\cpc #1 #2 #3 {{Comp. Phys. Commun.} {\bf#1} (#2) #3}
\def\sjnp #1 #2 #3 {{Sov. J. Nucl. Phys.} {\bf#1} (#2) #3}
\def\xx #1 #2 #3 {{\bf#1}, (#2) #3}
\def\hepph #1 {{\tt hep-ph/#1}}

\def\beq{\begin{equation}}
\def\beeq{\begin{eqnarray}}
\def\eeq{\end{equation}}
\def\eeeq{\end{eqnarray}}
\def\a0{\bar\alpha_0}

\def\b0{\beta_0}

\def\ee{e^+e^-}

\def\slashchar#1{\setbox0=\hbox{$#1$}           
     \dimen0=\wd0                                 
     \setbox1=\hbox{/} \dimen1=\wd1               
     \ifdim\dimen0>\dimen1                        
        \rlap{\hbox to \dimen0{\hfil/\hfil}}      
        #1                                        
     \else                                        
        \rlap{\hbox to \dimen1{\hfil$#1$\hfil}}   
        /                                         
     \fi}                                         %

\def\be{\begin{equation}}
\def\ee{\end{equation}}
\def\bea{\begin{eqnarray}}
\def\eea{\end{eqnarray}}
\def\lsim{\:\raisebox{-0.5ex}{$\stackrel{\textstyle<}{\sim}$}\:}

\def\slash{/\kern -5pt}

\def\ims #1 {\ensuremath{M^2_{[#1]}}}

\def\s22w{s_{2W}^2}

\title{Full One-loop Electro-Weak Corrections to Three-jet Observables at the 
$Z$ Pole and Beyond}
\preprint{
{CERN-PH-TH/2008-062 }\\
{LPT-ORSAY-07-130}\\
{SHEP-07-23}\\
{FNT/T-2008/02}\\
}
\author{C.M. Carloni-Calame$^{1,2}$, S. Moretti$^{2,3}$, F. Piccinini$^{4}$ and D.A. Ross$^{2,5}$\\
{$^1$ INFN, via E. Fermi 40, Frascati, Italy}
\\
{$^2$ School of Physics and Astronomy, University of Southampton}
{Highfield, Southampton SO17 1BJ, UK}
\\
{$^3$ Laboratoire de Physique Th\'eorique, Universit\'e Paris-Sud,
F-91405 Orsay Cedex, France}
\\
{$^4$ INFN - Sezione di Pavia, Via Bassi 6, 27100 Pavia, Italy}
\\
{$^5$ Theory Unit, Physics Department, CERN, Geneva 23, Switzerland}\\
E-mails: \email{c.carloni-calame@phys.soton.ac.uk}, 
\email{stefano@phys.soton.ac.uk}, \email{fulvio.piccinini@cern.ch}, 
\email{dar@phys.soton.ac.uk}}
\abstract{We describe the impact of the full one-loop EW  
terms of ${\cal O}(\alpha_{\rm S}\alpha_{\rm{EM}}^3)$ 
entering the electron-positron into 
three-jet cross-section from $\sqrt s=M_Z$ to TeV scale energies.
We include both factorisable and non-factorisable virtual corrections,
photon bremsstrahlung but not the real emission of $W^\pm$
and $Z$ bosons. Their importance for the measurement of $\alpha_{\rm S}$ 
from jet rates and shape variables is explained.}
\begin{document}
\section{Introduction}
\label{Sec:Intro}
Strong (QCD) and Electro-Weak (EW) interactions
are two fundamental forces of 
Nature, the latter in turn unifying
Electro-Magnetic (EM) and Weak (W) interactions
in the Standard Model (SM).
A clear hierarchy exists between the strength of these two interactions
at the energy scales probed by past and present high energy particle 
accelerators (e.g., LEP, SLC, HERA and Tevatron): 
QCD forces are stronger than EW ones. This is quantitatively
 manifest if one recalls that the value of the QCD coupling, 
$\alpha_{\mathrm{S}}$, measured at these machines is much larger 
than the EM one, $\alpha_{\rm{EM}}$, typically, by an order of
magnitude. This argument, however, is only valid in lowest order
in perturbation theory. 

A peculiar feature distinguishing QCD and EW effects in higher orders
is that the latter are enhanced by (Sudakov) double logarithmic factors,
$\ln^2(\frac{s}{M^2_{{W}}})$, 
which, unlike in the former, do not cancel for `infrared-safe' 
observables  \cite{Kuroda:1991wn,Beenakker:1993tt,Ciafaloni:1999xg,Denner:2000jv}. 
The origin of these `double logs' is well understood.
It is due to a lack of the Kinoshita-Lee-Nauenberg (KLN) 
\cite{KLN} type 
cancellations of Infra-Red (IR) -- both soft
and collinear -- virtual and real emission in
higher order contributions originating from $W^\pm$ (and, possibly,
$Z$: see Footnote 1) exchange. 
This is in turn a consequence of the 
violation of the Bloch-Nordsieck theorem \cite{BN} in non-Abelian theories
\cite{Ciafaloni:2000df}.
The problem is in principle present also in QCD. In practice, however, 
it has no observable consequences, because of the final averaging of the 
colour degrees of freedom of partons, forced by their confinement
into colourless hadrons. This does not occur in the EW case,
where the initial state has a non-Abelian charge,
dictated by the given collider beam configuration, such as in $e^+e^-$
collisions. 

These logarithmic corrections are finite (unlike in
QCD), as the masses of the weak gauge bosons provide a physical
cut-off for $W^\pm$ and $Z$ emission. Hence, for typical experimental
resolutions, softly and collinearly emitted weak bosons need not be included
in the production cross-section and one can restrict oneself to the 
calculation of weak effects originating from virtual corrections and
affecting a purely hadronic final state\footnote{By doing so, double logarithms 
also arise from virtual $Z$ exchange, despite there being no flavour change, simply
because KLN cancellations are spoilt by the remotion of real $Z$ emission.}.
Besides, these contributions can  be
isolated in a gauge-invariant manner from EM effects
\cite{Ciafaloni:1999xg}, 
at least in some specific cases, and 
therefore may or may not
be included in the calculation, depending on the observable being studied. 
As for purely EM effects,
since
 the (infinite) IR real photon emission cannot be resolved experimentally, 
this ought to be combined with the (also infinite) virtual one, through the 
same order, to recover a finite result, which is however not
doubly logarithmically enhanced (as QED is an Abelian theory).
 
In view of all this,  it becomes of crucial importance to assess
the quantitative relevance of such EW corrections
affecting, in particular, key QCD processes studied at past, present and 
future 
colliders, such as $e^+e^-\to3$~jets. 
 
\section{Three-jet Events at Leptonic Colliders}
\label{Sec:ee}
It is the aim of our paper to report on the computation of the full 
one-loop EW effects entering three-jet production in electron-positron
annihilation at any collider energy
via the subprocesses $e^+e^-\to\gamma^*,Z\to \bar 
qqg$\footnote{See Ref.~\cite{2jet} for the corresponding one-loop
corrections
to the Born process $e^+e^-\to q\bar q$ and Ref.~\cite{4jet} for the
$\sim n_{\rm f}$ component of those to $e^+e^-\to q\bar qgg$ (where 
$n_{\rm f}$ represents the number of light flavours).}.
Ref.~\cite{oldpapers} tackled part of these, in fact, restricted to the
case of $W^\pm$ and $Z$ (but not $\gamma$) exchange and 
when the higher order effects arise only from initial or final state
interactions 
(these represent the so-called `factorisable' corrections, i.e.,
those involving loops 
not connecting the initial leptons to the final quarks or gluons,
which are the dominant ones at $\sqrt s=M_Z$, where the width 
of the $Z$ resonance provides a natural cut-off for off-shellness
effects). The remainder, `non-factorisable' corrections,
while being typically small at $\sqrt s=M_{Z}$, 
are expected to play a quantitatively relevant role as $\sqrt s$ grows
larger.  Since, here, we study the full set
 the one-loop EW corrections, we improve on the
results of Ref.~\cite{oldpapers} in two respects: (i) we include now all
the non-factorisable terms; (ii) we also incorporate previously
neglected genuine QED corrections, including photon bremsstrahlung.
In contrast, we refrain here from computing $W^\pm$ and $Z$ boson 
bremsstrahlung (just like in \cite{oldpapers}), as we will argue that this may not enter the experimental
jet samples.
(For a study of the impact of the emission of real massive gauge bosons
in a variety of high energy processes, see Ref.~\cite{Baur-rad},
albeit in hadronic collision.)

Combining the aforementioned logarithmic enhancement associated with
the genuinely weak component of the EW corrections to the fact that
$\alpha_{\rm S}$ steadily decreases with energy, 
unlike 
$\alpha_{\rm{EW}}$,
in general, one expects 
one-loop EW effects to become comparable to QCD ones
 at future Linear Colliders
(LCs) \cite{LCs} running at TeV energy scales,
like those available at an International Linear Collider (ILC) or the
Compact LInear Collider 
(CLIC)\footnote{For example, 
at one-loop level,
in the case of the inclusive cross-section of $e^+e^-$
into hadrons, the QCD corrections are of  ${\cal O}
(\frac{\alpha_{\mathrm{S}}}{\pi})$, whereas
the EW ones are of ${\cal O}(\frac{\alpha_{\mathrm{EW}}}{4\pi}\ln^2
\frac{s}{M^2_{{W}}})$, where $s$ is the collider centre-of-mass
(CM) energy
squared, so that at $\sqrt s\approx1.5$ TeV the former are identical to the latter,
of order 9\% or so.}. In contrast, 
at the $Z$ mass peak, where logarithmic
enhancements are not effective, one-loop EW corrections are expected to appear
at the percent level, hence being of limited relevance at
LEP1 and SLC, where the final error on $\alpha_{\mathrm{S}}$
is of the same order or larger \cite{Dissertori}, but of crucial importance
at a GigaZ stage of a future LC \cite{oldpapers}, where the relative accuracy
of $\alpha_{\mathrm{S}}$ measurements is expected to be at the
$0.1\%$ level or better~\cite{Winter}.
On the subject of higher order QCD
effects, it should be mentioned here that a great deal of effort has    
recently been devoted to evaluate two-loop contributions
to the three-jet process \cite{QCD2Loops}
while the one-loop QCD results have been known for quite some time \cite{ERT}.

As intimated, in the case of $e^+e^-$ annihilations, the most important QCD quantity to be 
extracted from multi-jet events is $\alpha_{\mathrm{S}}$.
The confrontation of the measured value of the strong coupling
constant with that predicted by the theory through the 
renormalisation group evolution is an important test of the 
SM. Alternatively, it may be an indication of new physics, when its typical mass scale is 
larger than the collider energy, so that the new particles cannot be
produced as `real' detectable states
but may manifest themselves through `virtual' effects. 
Not only jet rates,
but also
jet shape observables would be affected.

The calculation we are presenting here involves the full
one-loop EW corrections
to three-jet observables in electron-positron annihilations
generated via the interference of the graphs
in Figs.~\ref{fig:selfenergiesGB}--\ref{fig:pentagons} with the tree-level 
ones for $e^+e^-\to \gamma^*,Z\to \bar q qg$.
%
\FIGURE[t]{
\includegraphics[width=12cm]{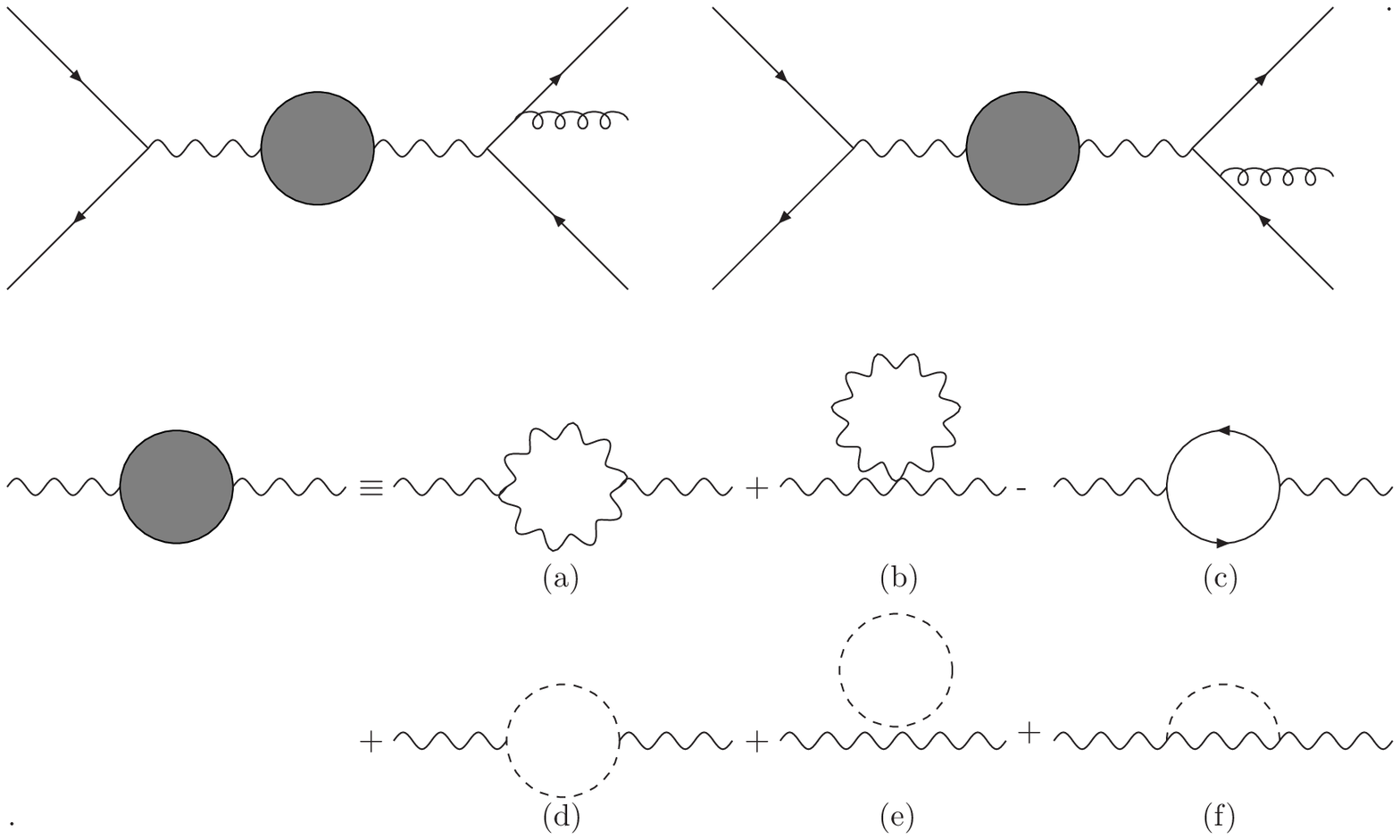}
\caption{Internal and external bosonic self-energy graphs.
The shaded blob on the 
wavy lines represents all the contributions to the gauge boson 
self-energy and is dependent on the Higgs mass, as detailed in the graphical equation. 
These corrections apply to the photon self-energy, the $Z$ boson
self-energy and the photon-$Z$ mixing.
The  gauge bosons inside the loops
are $W^\pm$ and the  scalars are charged
Goldstone bosons. In the case of the self-energy of the $Z$, in
graphs  (d)--(f), the  gauge boson inside the loop
 can be a $Z$ and the  scalar a 
Higgs boson.}
\label{fig:selfenergiesGB}}
%
\FIGURE[t]{
\includegraphics[width=11cm]{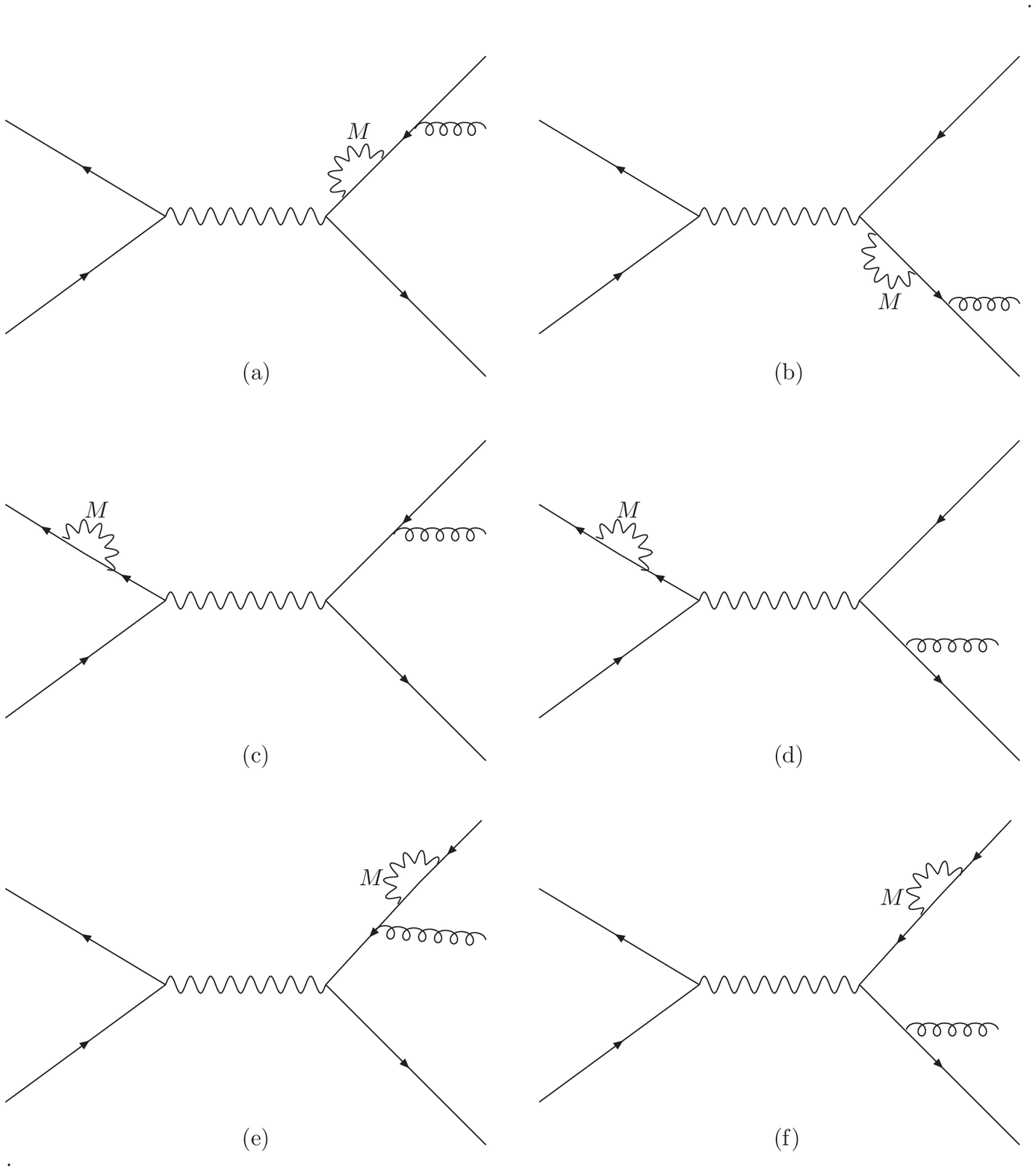}
\caption{Internal and external fermionic self-energy graphs.
The gauge boson mass, $M$, can take the values $\lambda, \ M_Z$
or $M_W$. For the external self-energies, we display only the self-energy
insertion on the incoming positron (c), (d) or outgoing antiquark (e), (f).
There is an identical contribution from self-energy insertions on the
electron or quark, but each carries a combinatorial factor of $\frac{1}{2}$.}
\label{fig:selfenergiesF}}
\FIGURE[t]{
\includegraphics[width=11cm]{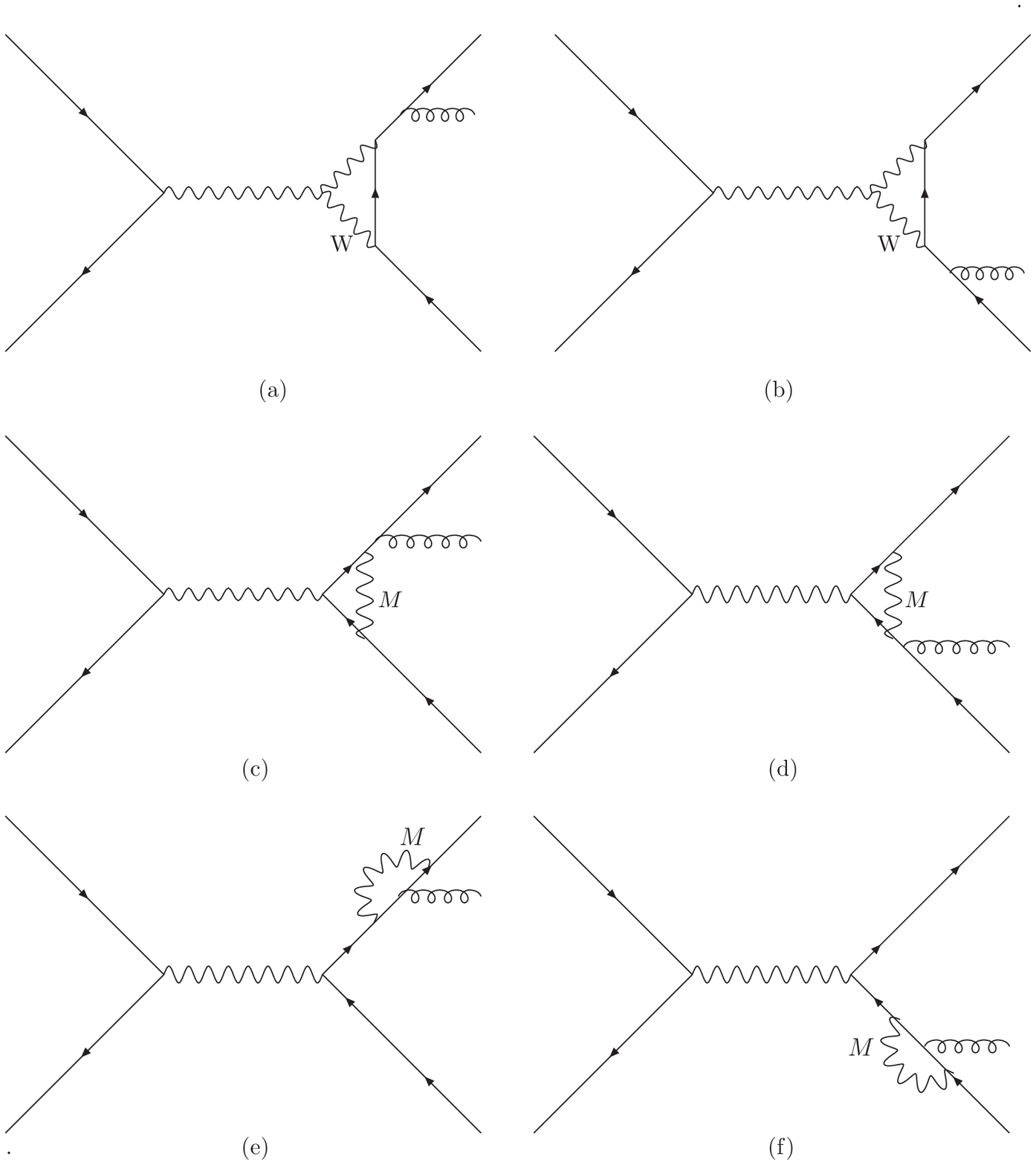}
\caption{Vertex graphs. Corrections on the (anti)quark line.  
In graphs (a) and (b) the gauge boson mass inside the loop must be $M_W$
whereas for graphs (c) to (f) the gauge boson mass
 $M$ can take the values $\lambda, \ M_Z$
or $M_W$.}
\label{fig:vertices}}
Hence, our calculation not
only accounts for the aforementioned double logarithms, but also all single ones
as well as the finite terms arising through the complete 
 ${\cal O}(\alpha_{\rm S}\alpha_{\rm{EW}}^3)$. We will account for all
possible flavours of (anti)quarks in the final state, with the exception
of the top quark. The latter however appears in some of the loops whenever
a $b\bar bg$ final state is considered. 
\FIGURE[t]{
\includegraphics[width=11cm]{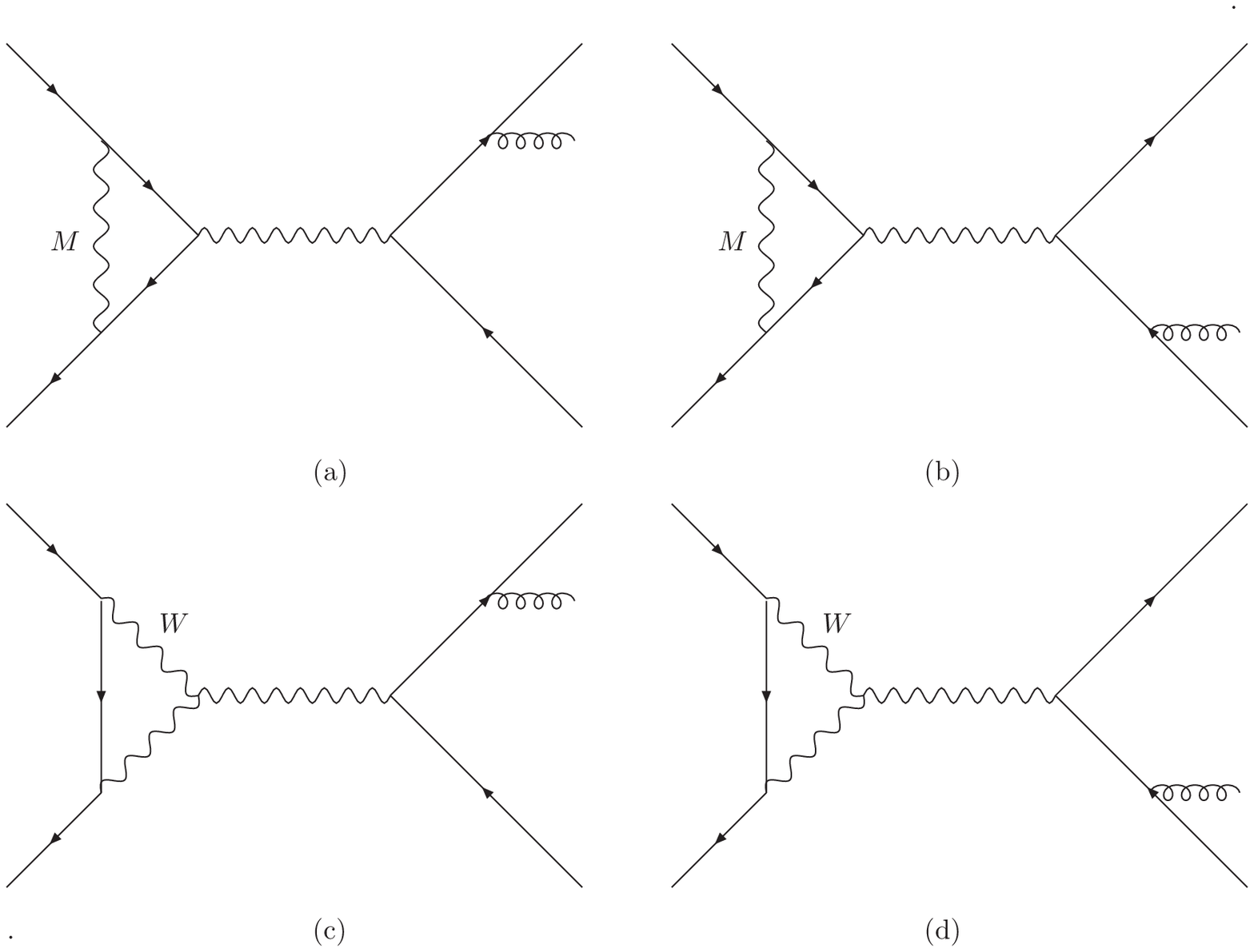}
\caption{Vertex graphs. Corrections on the electron/positron line.  
In graphs (a) and  (b) the gauge boson mass
mass $M$ can take the values $\lambda, \ M_Z$
or $M_W$ whereas in graphs (c) and (d)
 inside the loop there must be $M_W$.}
\label{fig:eevertices}}
\FIGURE[t]{
\includegraphics[width=11cm]{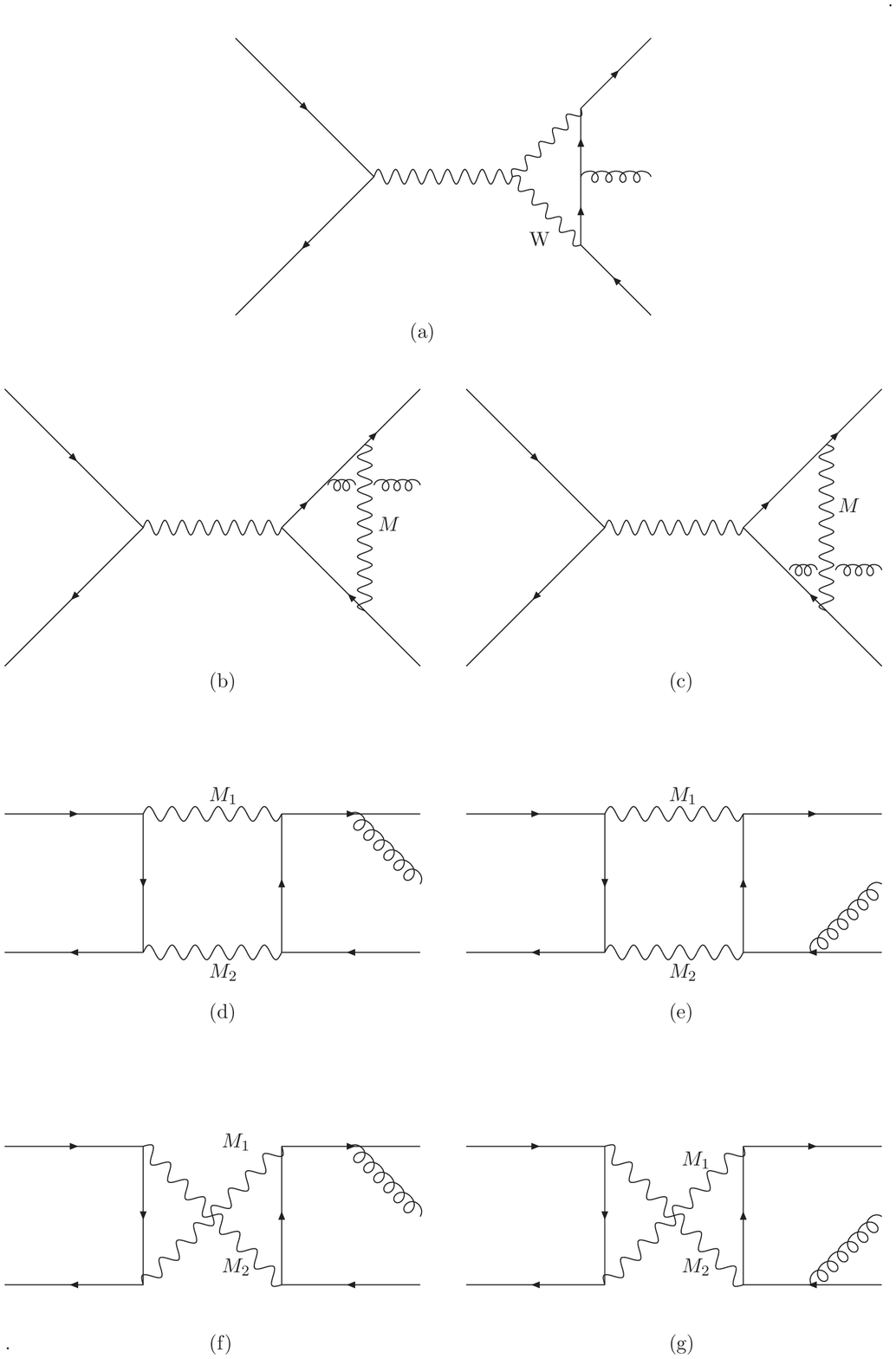}
\caption{Box graphs. In the non-factorising corrections (d)--(f),
the gauge boson masses $M_1, \, M_2$ can each take the values
$\lambda$ or $M_Z$ or they can both be $M_W$.}
\label{fig:boxes}}
\FIGURE[t]{
\includegraphics[width=14cm]{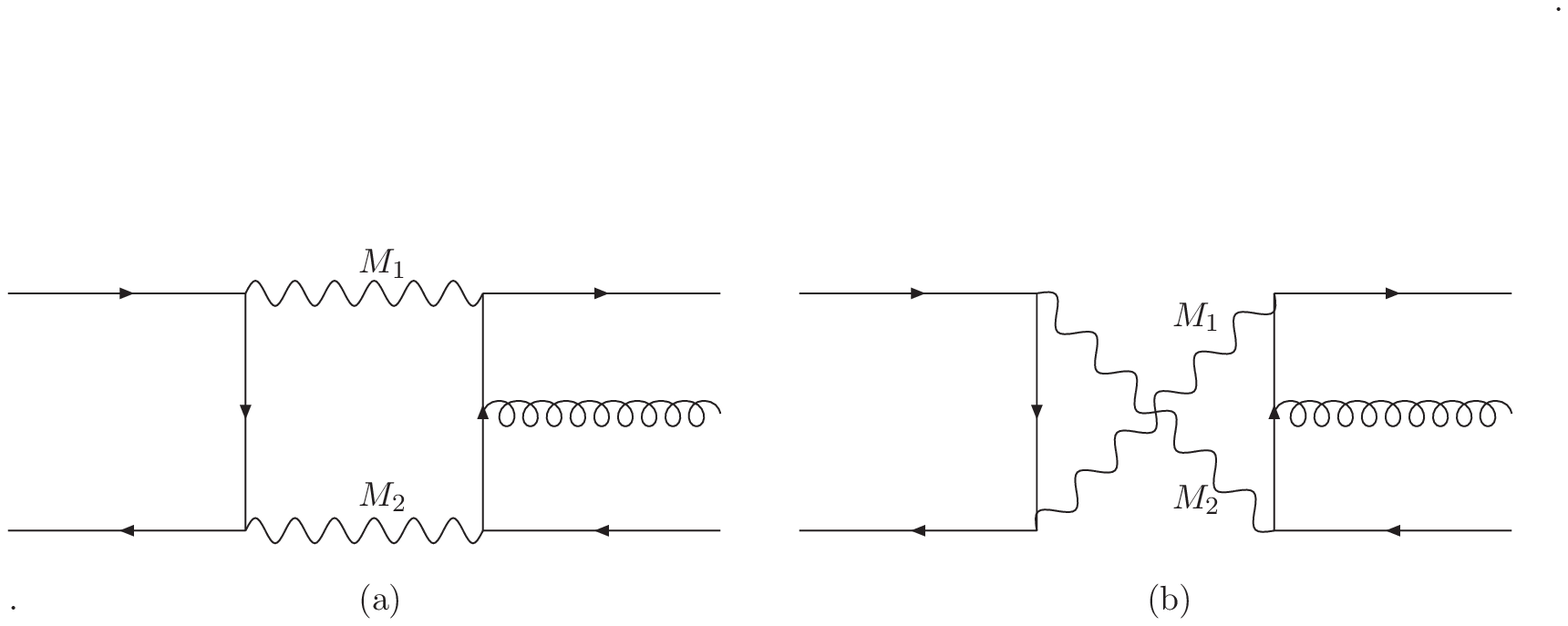}
\caption{Pentagon graphs. The gauge boson masses can take the same
values as in the case of the non-factorising box graphs.}
\label{fig:pentagons}}

We will show that all such corrections can range from a few percent 
to a few tens of percent, both at $\sqrt s=M_Z$ and LC energies,
depending on the observable under study, with the QED component
being preponderant at low energy and the weak one emerging more and more
as the latter increases. Altogether,  
while their impact is not dramatic in the context
of LEP1 and SLC physics at a GigaZ stage of future LCs
they ought to be taken into account in the experimental fits.
This is even more the case of future LCs running at and beyond
the TeV range\footnote{Another relevant phenomenological aspect of our
calculation would pertain
to the case in which the bremsstrahlung photon is resolved, so that
a three-jet plus one-photon sample can be defined and compared to a 
four-jet sample in order to extract genuine non-Abelian QCD effects,
as detailed in \cite{3j1a} -- and references  therein --
but we postpone this analysis to a future work.}.

The plan of the rest of the paper is as follows. In the next Section,
we describe the calculation. Then, in Sect.~\ref{Sec:Results},
we present some numerical results (a preliminary account of which was given in
Ref.~\cite{RADCOR}). We conclude in Sect.~\ref{Sec:Conclusions}.

\section{Calculation}
\label{Sec:Calculation}

In Ref.~\cite{Maina:2002wz},
  which concentrated only on the factorisable
corrections to three-jet production (i.e., neglecting any interaction between
the incoming electron-positron pair and the final-state (anti)quarks), 
the method of 
helicity Matrix Elements (MEs) was adopted. In the present case such
an approach is not convenient because we have to address the problem of IR 
divergent contributions from loops in which a photon
is exchanged. As intimated already, this means that we also need to consider 
the real photon (bremsstrahlung)
contributions, summed over the helicities of the emitted photon. 
Moreover, since in our numerical simulations, as we 
will see later, we regularise the collinear singularities with finite 
fermion masses, 
the introduction of these terms would complicate the expressions 
obtained within the helicity formalism.
For the virtual corrections, we therefore adopt
the more traditional approach of considering all possible interferences 
between the one-loop and 
tree-level graphs\footnote{We have however verified that, limitedly
to the case of factorisable corrections, we can reproduce the (unpolarised)
results of  Ref.~\cite{oldpapers}.}.
We have organized the former into ``prototype graphs'', shown in 
 Figs.~\ref{fig:selfenergiesGB}--\ref{fig:pentagons}, which were calculated as 
general functions of couplings and internal masses, which appear 
inside 
Passarino-Veltman (PV) functions
\cite{PV}, which are then determined numerically with the appropriate 
couplings and internal masses. In doing so, we have made extensive use of the
{\tt FORM}~\cite{FORM} program.

In anticipation of an electron-positron collider in which the incoming beams 
can be polarised, we have however inserted a helicity projection operator 
into the electron line and obtained
separate results for left-handed and right-handed incoming
electrons\footnote{As we are taking massless incoming fermions,
 the helicity of the positron
is simply the opposite to that of the electron.}. 
For genuinely weak interaction
corrections, this is of particular interest, since such corrections violate 
parity conservation.
Unfortunately, this occurs already at tree-level,
 owing to the contribution from 
exchange of a $Z$ boson,
but these higher order 
corrections are also peculiarly dependent on the incoming lepton helicity 
and thus one 
would expect the two parity-violating effects be 
distinguishable after the collection of sufficient events. (We will
devote a separate paper to the study of polarised incoming beams.)

In this connection, it is worth mentioning that as we are dealing with weak 
interactions, which involve axial couplings,
and as we furthermore wish to distinguish between left- and right-helicity
incoming electrons, the only unambiguous way to handle loop integrals is to 
use the `modified' Dimensional Reduction
($\overline{\mathrm{DR}}$)
as opposed to the `modified' Minimal Subtraction
($\overline{\mathrm{MS}}$) renormalisation prescription.
(We use 
$\mu=M_Z$ 
as the subtraction scale of ultraviolet divergences.)
It is therefore necessary
to determine the value of the fine-structure constant $\alpha_{\rm{EM}}$ in that scheme 
at the subtraction point $\mu=M_Z$\footnote{We emphasise that we are 
{\it not} using the complex mass subtraction
scheme of Ref.~\cite{complexmass}, but rather the $\overline{\mathrm{DR}}$ one
in which all counterterms are taken to be real. For the internal gauge boson 
self-energies
the mass subtraction is the on-shell self-energy. Provided this is taken
in conjunction with the corresponding values for the EM coupling,
$\alpha_{\rm{EM}}$, and the
weak mixing angle, $\theta_W$ (as discussed here), this is an equally 
consistent scheme
in which the imaginary parts of the renormalised couplings in the scheme 
of~\cite{complexmass}
are reproduced by the imaginary parts of the subtracted 
gauge boson self-energies.}. 
For the light fermion (including the $b$-quark) contributions
we can integrate the $\beta$-function through the thresholds at
$Q^2=4m_i^2$, for all charged fermions with mass $2m_i \, < \, M_Z$.
There is a difficulty here with the light quarks as one expects substantial QCD
corrections and the mass thresholds are well below the regime where perturbative QCD 
is reliable. This problem is usually addressed by considering  the total cross-section
for electron-positron annihilation in the resonant region and applying a finite energy sum rule.
This has been carried out in detail in Ref.~\cite{pivovarov}. The upshot of this numerical analysis is that the
effect of all light fermions can be simulated by using ordinary perturbation theory with a threshold of
15 MeV for the $u$- and $d$-quarks and 1 GeV for the $s$-quark, yielding a value
$\alpha_{\rm{EM}}^{\mathrm{MOM}}(M_Z^2)=1/128.2$, in the MOMentum (MOM) subtraction renormalisation scheme. 
However, there is a finite difference between this 
prescription and a genuine $\overline{\mathrm{DR}}$ determination
arising from:
\begin{enumerate}
\item contributions to the photon self-energy from loops of $W^\pm$ gauge bosons (and their attendant Goldstone bosons);  
\item a contribution from the non-vanishing $Z$ photon mixing part of the self-energy at zero momentum;
\item a contribution from the loop correction to the EM vertex involving internal $W^\pm$'s, which does {\it not}
cancel by virtue of the QED Ward identity, owing to the non-Abelian nature of the coupling of the photon to 
charged gauge bosons.
\end{enumerate}
These corrections have been considered in Ref.~\cite{VG} in the
$\overline{\mathrm{MS}}$ scheme. In the $\overline{\mathrm{DR}}$ scheme
we find that these give a total contribution of 
\begin{equation}
\Delta\left(\alpha_{\rm{EM}}^{\overline{\mathrm{DR}}}(M_Z^2)\right) \ = \ - \frac{7\alpha^2_{\rm{EM}}(M_Z^2)}{4\pi}
 \ln\left(\frac{M_W^2}{M_Z^2}\right),
\end{equation}
where $s_W$ and $ c_W$ are $\sin\theta_W$ and  $\cos\theta_W$, respectively.
This introduces a negligible correction.

Finally, we need to account for the difference between the MOM and the $\overline{\mathrm{DR}}$ schemes,
from the light fermions (including the $b$-quark), and a contribution from the $t$-quark, which
 lead to the more substantial corrections
\begin{equation}
\Delta \left( \alpha_{\rm{EM}}^{\overline{\mathrm{DR}}}(M_Z^2) \right)
\  = \  
  \frac{\alpha^2_{\rm{EM}}(M_Z^2)}{3\pi}
 \left[ \sum_i q_i^2 \ln(4)
 -\frac{4}{3}  \ln\left(\frac{m_t^2}{M_Z^2}\right)  \right], \end{equation}
where the sum is over all fermions whose electric charge is $q_i \, e$, except the $t$-quark. 
This gives us a final value of
\begin{equation} 
 \alpha_{\rm{EM}}^{\overline{\mathrm{DR}}}(M_Z^2) \ = \ \frac{1}{127.7}. \end{equation}

Loops containing one or two photons give rise to IR divergences. For all 
of these divergent
 loop integrals we have a means of isolating the IR part in the 
$\overline{\mathrm{DR}}$
scheme in $4-2\epsilon$ dimensions. This is explained in detail
in the appendix. For {\it all} of these IR divergent integrals we have 
checked numerically that
the introduction of a common mass, $\lambda$, for {\it both} the internal fermion 
lines and the internal
 photon lines reproduces these integrals upon the replacements
$$ \pi^{-\epsilon} \Gamma(\epsilon) 
 \ \to \ 
\ln(\lambda^2),
 $$
 $$\pi^{-\epsilon}\frac{\Gamma(\epsilon)}{\epsilon}
 \ \to \ \frac{1}{2} \ln^2(\lambda^2), $$
provided $\lambda$ is chosen sufficiently small, but not so small as to 
introduce numerical instabilities.
In each case we have identified a substantial plateau region in 
$\lambda$ where the two methods of calculation agree
to a very high degree of accuracy. 

Having established that this works, we then abandon the use of dimensional 
regularisation for the case
of IR divergences and simply insert a mass regulator $\lambda$ for the photon 
and an independent mass $m_f$ for all
fermions. This is also done in the case of the bremsstrahlung contribution 
before integrating over the 
phase space for the emitted photon. The reason why the mass 
regularisation of IR (i.e., soft/collinear) singularities works is linked to the 
fact that the tree-level diagrams only involve neutral currents, so that 
the EM and purely weak corrections are separately gauge invariant, as mass regularisation is
well known to work in QED.  
By varying both of these parameters and 
noting the insensitivity 
of the final results to these changes, we 
have checked the expected cancellations
of these divergences between the virtual correction and real emission. 

A new feature of this calculation, which was not present in the case
of factorisable corrections only dealt with in \cite{Maina:2002wz}, is the
occurrence of pentagon graphs, as shown in Fig.~\ref{fig:pentagons}, which arise 
from interactions with two gauge bosons exchanged
in the $s$-channel and the gluon  emitted from an internal (anti)quark. Such graphs 
involve five-point PV functions 
with up to three powers of momenta in the numerator. We have handled these in two 
separate ways (with two independently developed codes), in order to check
for possible numerical instabilities. In the first case the integrals
are simply evaluated using routines in {\tt LoopTools v2.2}~\cite{looptools},
for which these tensor-type pentagon integrals are now implemented according to 
the reduction formalism of Ref.~\cite{DD}\footnote{We implemented also 
directly in an independent {\tt FORTRAN} routine the expressions for the 
five-point functions of Ref.~\cite{DD}, 
finding agreement with {\tt LoopTools} up to available digits
precision.}. In the other we use the standard PV
reduction to express scalar products of the loop momentum $k$ with
external momenta $p_i$ in terms of the denominators of
propagators adjacent to the $i$-{th} vertex. This reduction is carried out 
exhaustively until only scalar pentagon
integrals appear. The latter are available in the library {\tt FF1.9}~\cite{FF1.9}. 
A comparison of  the numerical results provided by the two codes for 
a sample of phase space points yielded satisfactory agreement between the two 
methods.

The square MEs and the interferences for the bremsstrahlung process 
have been calculated directly using 
{\tt ALPHA}~\cite{ALPHA}
and checked against both {\tt MadGraph}~\cite{SL}
and the results based on the codes of Ref.~\cite{BMM}, 
finding perfect agreement between all these implementations.

Regarding the integration over the photon phase space, this has 
been split into two regions
\begin{enumerate}
\item $E_\gamma \ \leq \ \Delta E $.
\item $E_\gamma \ > \ \Delta E $.
\end{enumerate}
For the first region, the MEs have been approximated using the eikonal 
approximation in which a fermion
with momentum $p$  emitting a photon (with polarisation $\epsilon$) 
has an associated 
vertex $2p \cdot \epsilon$ and 
the photon momentum is neglected in all numerators. Here, the phase space integration 
has been performed 
analytically
assuming a photon mass $\lambda$. 
In the second region, the full MEs are employed but the photon mass is set 
to zero. 
Here, we integrate over {\it all} the phase space available to the photon. 
The integrations for both the three- and four-particle  final states 
were performed numerically with Monte Carlo (MC) techniques, so that
any desired cut on the final state can be imposed.

The three- and four-particle Lorentz invariant phase space reads as
\begin{eqnarray}
d^5\Phi_3&=&\frac{1}{8(2\pi)^5}
\frac{d^3\mathbf{p}_{q}}{E_q}
\frac{d^3\mathbf{p}_{\bar{q}}}{E_{\bar{q}}}
\frac{d^3\mathbf{p}_{g}}{E_{g}}
\,\, \delta^4(p_{e^+}+p_{e^-} - p_{q} - p_{\bar{q}}-p_g),
\nonumber\\
d^8\Phi_4&=&\frac{1}{16(2\pi)^8}
\frac{d^3\mathbf{p}_{q}}{E_q}
\frac{d^3\mathbf{p}_{\bar{q}}}{E_{\bar{q}}}
\frac{d^3\mathbf{p}_{g}}{E_{g}}
\frac{d^3\mathbf{p}_{\gamma}}{E_{\gamma}}
\,\, \delta^4(p_{e^+}+p_{e^-} - p_{q} - p_{\bar{q}}-p_g-p_\gamma),
\label{34phasespace}
\end{eqnarray}
respectively.
We choose as five (eight) independent integration variables for the
three (four) body phase space the angles of the quark, the energy and
the angles of the gluon (and the energy and the angles of the
photon). With this choice, Eq.~(\ref{34phasespace}) can be written as 

\begin{eqnarray}
d^5\Phi_3&=&\frac{1}{8(2\pi)^5}
\frac{|\mathbf{p_q}| E_g}{|E_{\bar{q}}+E_q\left(1+
\mathbf{p}_g\cdot\mathbf{p}_q/|\mathbf{p}_q|^2\right)|}
dE_g d^2\Omega_g d^2\Omega_q,
\nonumber\\
d^8\Phi_4&=&\frac{1}{16(2\pi)^8}
\frac{|\mathbf{p_q}| E_g E_\gamma}{|E_{\bar{q}}+E_q\left[1+
\left(\mathbf{p}_g + \mathbf{p}_\gamma\right)\cdot\mathbf{p}_q/|\mathbf{p}_q|^2\right]|}
dE_gd^2\Omega_gdE_\gamma d^2\Omega_\gamma d^2\Omega_q.
\label{34phasespacereduced}
\end{eqnarray}
In Eq.~(\ref{34phasespacereduced}), all the kinematical quantities are
derived from the independent variables by imposing four-momentum
conservation and mass-shell relations and are calculated in the
rest frame of the incoming $e^+e^-$ system.

In order to perform the numerical integration over the phase space, the
independent variables are sampled according to the peaking structure
of the integrand function. Standard multi-channel importance sampling
techniques are used. Concerning the importance sampling of the three-
and four-body phase space collinear configurations, our treatment
closely follows the multi-channel approach described in appendix 
{A.3} of Ref.~\cite{bbatnlo}.
\TABLE[t]{\parbox[h]{\textwidth}{
\begin{center}
\begin{tabular}{| c | c | c |}
\hline
 $\lambda^2$ (GeV$^2$) & $5\cdot 10^{-10}$ & $1\cdot 10^{-13}$\\
\hline
 $\sigma_{V}+\sigma_{S}$ (pb), $\sqrt{s}=90$ GeV&  $-4181.6\pm 5$&
 $-4181.2\pm 5$\\
 $\sigma_{V}+\sigma_{S}$ (pb), $\sqrt{s}=350$ GeV&  $-1.2293\pm 0.005$ &  $-1.2293\pm 0.005$\\
\hline
\end{tabular}
\caption{
Variation of the soft plus virtual cross section 
$\sigma_V+\sigma_{S}$ for two different values of $\lambda$ for
  $\sqrt{s}=90$ and $350$ GeV. Here, only
order $\alpha_{\rm{EM}}$ QED contributions are included, the cuts are
specified in the text and $2\Delta E/\sqrt{s}$ is $10^{-4}$. The sum
over the final-state quark flavours is taken.}
\label{tab:lambda}
\end{center}}}

As intimated, we have checked that the 
final result is insensitive to variations of (small values) 
of not only the fictitious photon mass $\lambda$ but also the 
separator $\Delta E$. Table~\ref{tab:lambda} shows the sum of the 
virtual, $\sigma_V$, and soft, $\sigma_S$, cross section
(integrated on $E_\gamma$ from $\lambda$ to $\Delta E$), while
spanning a wide range of values for $\lambda$.
Table~\ref{tab:deltae} shows the real radiation cross section, 
$\sigma_{\rm{real}}$, where the integration on $E_\gamma$ is split in
the soft part $\lambda \leq E_\gamma \leq \Delta E$ and the hard part
$\Delta E \leq E_\gamma \leq E_\gamma^{\rm{max}}$. (Parameters and cuts 
on the hadronic system are defined in the following Section.) 

When the real photon emission process is combined with the virtual 
corrections we have checked that the limit
$\lambda \to 0$ may be taken, whilst keeping the fermion mass $m_f$ small but 
non-zero. The collinear divergence
obtained when $m_f \to 0$ cancels with the well-known exception of an overall 
large logarithm ($\ln(s/m_f^2)$),
which is associated with the Initial State Radiation (ISR) induced by
the incoming electrons and positrons.
\TABLE[t]{\parbox[h]{\textwidth}{
\begin{center}
\begin{tabular}{| c | c | c |}
\hline
 $2 \Delta E/\sqrt{s}$ & $10^{-3}$ & $10^{-4}$\\
\hline
 $\sigma_{\rm{real}}$ (pb), $\sqrt{s}=90$ GeV & $5054.\pm 1.$  &
$5050.\pm 2$   \\
\hline
 $\sigma_{\rm{real}}$ (pb), $\sqrt{s}=350$ GeV & $1.978\pm 0.001 $  &
$1.979\pm 0.001 $ \\
\hline
\end{tabular}
\caption{\label{tab:deltae}
Variation of the four-body cross section 
$\sigma_{\rm{real}}$ for two different values of $\Delta E$, wherein 
$\lambda^2$ is fixed to the value $10^{-13}$ GeV$^2$.
Again, only order $\alpha_{\rm{EM}}$
 QED contributions are included, $\sqrt{s}$ is $90$ and $350$ GeV, the cuts are
specified in the text and the sum over the final-state quark flavours
is taken.
}
\end{center}
}}

It is well known that, in the case of QCD, such a remnant 
collinear divergence is absorbed into the momentum dependence of the 
Parton Distribution Functions (PDFs).  
In the case of
electron-positron colliders this large correction is always present, 
but it is universal to all processes and would therefore tend to mask the 
purely weak corrections. However, for sensible numerical 
results, it has to be accounted for to all orders of perturbation 
theory, e.g., within the so-called electron/positron structure function formalism~\cite{sf}, 
which automatically resums in QED all Leading Logarithmic (LL) terms. 
In Ref.~\cite{MNP} a method of combining consistently 
resummed LL calculations with exact ${\cal O}(\alpha_{\rm{EM}})$ ones 
has been devised both in additive and factorisable form. 
Here, we will adopt the additive approach, 
which amounts to write the differential cross section as 
\begin{equation}
d\sigma = d\sigma_{\rm LL} - d\sigma_{\rm LL}^{\alpha} + d\sigma_{\rm{exact}}^{\alpha},
\label{eq:comb}
\end{equation}
where $d\sigma_{\rm LL}$ and $d\sigma_{\rm LL}^\alpha$ are defined by
\begin{eqnarray}
d\sigma_{\rm LL}&=& \int dx_1dx_2D(x_1,s)D(x_2,s)d\sigma_{0}(x_1x_2s),\nonumber\\
d\sigma^\alpha_{\rm LL}&=& \int dx_1dx_2[D(x_1,s)D(x_2,s)]_{\alpha}
d\sigma_{0}(x_1x_2s)
\label{LLandLLa}
\end{eqnarray}
and where $D(x,Q^2)$ is the electron structure function whilst
$[D(x_1,s)D(x_2,s)]_{\alpha}$ is the order $\alpha_{\rm{EM}}$ term of
$D(x_1,s)D(x_2,s)$. In Eq.~(\ref{eq:comb}), the order $\alpha_{\rm{EM}}$ part
of the up-to-all-order resummed cross section in LL approximation 
is replaced by the exact order $\alpha_{\rm{EM}}$ correction or, in other words,
the higher-order (beyond order $\alpha_{\rm{EM}}$) LL corrections are added to
the exact order $\alpha_{\rm{EM}}$ calculation.

There remains one kinematic region where our results would
 not be numerically reliable. 
In the case of real photon emission, the aforementioned
ISR can lead to the situation in which the remaining 
sub-energy of the incoming electron-positron pair
is close the the $Z$ mass and hence close to the pole of the propagator in 
the case of $Z$ boson $s$-channel exchange. In order to avoid this
the width, $\Gamma_Z$, of the $Z$ boson has been included in the propagator. 
For consistency, this means that the same width has to 
be included in the $Z$ propagator for the virtual corrections. 
This can be done in a consistent way in the $\overline{\mathrm{DR}}$ scheme
used in this paper, in which renormalised couplings
remain real,  as well as the  the complex mass 
scheme~\cite{complexmass}, where the $W$ and $Z$ masses are defined 
as the locations of the propagator poles in the complex plane, and 
the couplings become complex. The essential ingredient for the 
evaluation of virtual corrections is the ability to compute 
one-loop integrals with complex internal masses. The package 
{\tt LoopTools v2.2}~\cite{looptools} implements complex masses 
only for two-point, three-point and IR-singular four-point 
functions. We implemented in {\tt LoopTools} the general expression 
for the scalar four-point function of Ref.~\cite{tHV}, 
valid also for complex masses. Particular attention has been 
devoted to the occurrence of numerical instabilities in 
certain regions of phase space because of strong cancellations. 
The new routine for the scalar four-point 
function with complex masses has been tested with extended 
{\tt complex*32} precision, finding perfect numerical agreement. 
Actually, for the present study, the implementation of the complex 
mass scheme is only partial since we have kept real couplings and thus 
our scheme is equivalent to the fixed width scheme~\cite{fixedwidth} 
implemented at the full one-loop level. In principle this could produce 
a bad high energy behaviour of the corrections. We tested the stability 
of our predictions at high energy by switching off the $W/Z$ width 
in the virtual and real corrections, checking that the effect of the
width is smaller and smaller as the energy raises up to 1 TeV: 
the differences on the pure virtual corrections with and without $W/Z$ width 
(and setting $\lambda^2=5\cdot 10^{-10}$~GeV$^2$) are $0.5\%$ at
$\sqrt{s}=$350~GeV, $0.2\%$ at $\sqrt{s}=$700~GeV and $0.1\%$ at
$\sqrt{s}=$1~TeV, in units of the Born cross section.

We have neglected the masses of light quarks throughout. However, 
in the case in which the final state
contains a $b\bar b$ pair, whenever there is a $W^\pm$ boson
in the virtual loops, account had to be taken of the mass of the 
top (anti)quark. Furthermore, in such cases, it was also
necessary to supplement the graphs shown in  
Figs.~\ref{fig:selfenergiesGB}--\ref{fig:pentagons}
with the corresponding diagrams in which one or two internal $W^\pm$ 
bosons are
 replaced by charged Goldstone bosons, as appropriate. Having done 
all this, 
we are therefore in a position to present the results for such 
`$b$-jets' separately, though this will also be done in another publication.

\section{Numerical Results}
\label{Sec:Results}

Before proceeding to show our results, we should mention the
parameters we set for our simulations. We have
taken the top (anti)quark to have a
mass $m_t = 171.6$ GeV. The $Z$ mass used was $M_Z = 91.18$ GeV and was
related to the $W^\pm$ mass, $M_W$, via the SM formula
$M_W = M_Z \cos \theta_W$, where $\sin^2 \theta_W =$ 0.222478. The $Z$
width was $\Gamma_Z = 2.5$ GeV.
Also notice that, where relevant, Higgs contributions are included
with $M_H=115$ GeV.
For the strong
coupling constant, $\alpha_{\rm{S}}$, we have used the two-loop expression with
$\Lambda^{(nf=4)}_{\rm{QCD}}=0.325$ GeV in the
$\overline{\mathrm{MS}}$ scheme, yielding
$\alpha_{\rm{S}}^{\overline{\mathrm{MS}}}(M_Z^2)=0.118$.

All the numerical results presented in this section are obtained
considering a realistic
experimental setup. Namely, partonic momenta are clustered into jets 
according to the
Cambridge jet algorithm~\cite{cambridge} (e.g., when $y_{ij}<y_{cut}$
with $y_{cut}=0.001$), the jets are required to lie in the central
detector region $30^\circ<\theta_{\mathrm{jets}}<150^\circ$ and we request that
the invariant mass of the jet system $M_{3j}$ is larger than
$0.75\times\sqrt{s}$.
If a real photon is present in the final state, it is
clustered according to the same algorithm, but we require that
at least three ``hadronic'' jets are left at the end
(i.e., events in which the photon is resolved are rejected). 
Notice that this procedure  serves a twofold purpose. On the
one hand, from the experimental viewpoint, a resolved (energetic and isolated)
single photon is never treated as a jet. On the other hand, from a theoretical
viewpoint, this enables us to remove divergent contributions appearing 
whenever an unresolved real gluon is produced via an IR
 emission, as
we are not computing here ${\cal O}(\alpha_{\rm S}\alpha_{\rm{EM}}^3)$
one-loop QCD corrections to $e^+e^-\to q\bar q\gamma$. Finally, we sum
over the final-state quarks. In order to show the behaviour of the
corrections we are calculating, other than
scanning in the collider energy, we have considered here the three discrete
values of $\sqrt{s}=M_Z$ GeV, $\sqrt{s}=350$ GeV and $\sqrt{s}=1$ TeV.

\FIGURE[t]{
\includegraphics[width=12cm]{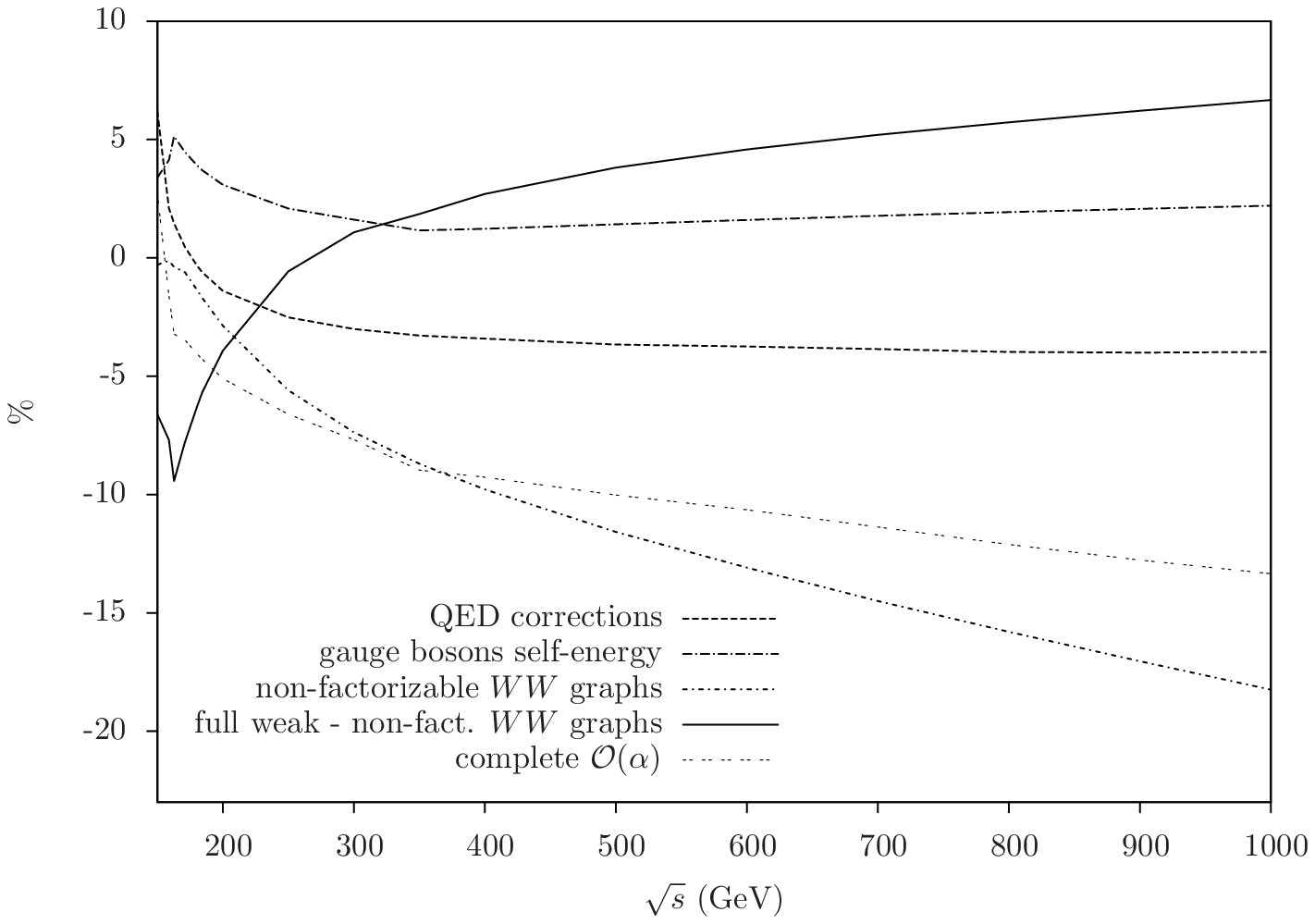}
\caption{Relative effect on the integrated cross section due to
  different contributions to the order $\alpha\equiv
\alpha_{\rm{EM}}$ correction, as a
  function of the CM energy. For setup and input
  parameters, see the text.}
\label{energyscan}}
In Fig.~\ref{energyscan}, the relative effects\footnote{The effects are
relative to the lowest-order cross section.} on the cross
section (integrated within the experimental cuts defined above)
induced by different contributions to the order $\alpha_{\rm S}
\alpha_{\rm{EM}}^3$
correction are plotted as a function of the CM energy, in
the range from 150 GeV to 1 TeV. The curves represent the effect of
the QED (virtual and real) corrections only, the effect of the gauge
bosons self-energy corrections (given by the graphs of
Fig.~\ref{fig:selfenergiesGB}), the
effect of the non-factorisable graphs of Figs.~\ref{fig:boxes}
(d)--(f) and~\ref{fig:pentagons} with $WW$
exchange\footnote{This is a gauge invariant subset of the complete 
correction.}, the effect of
the weak corrections with the non-factorizing $WW$ graphs removed
(labelled as ``full weak - non-fact $WW$ graphs'') 
 and the total effect as the sum of the previous
ones: the total effect is increasingly negative, reaching the $-13\%$ level at
1 TeV. It is worth mentioning that, as far as the non-factorisable
$WW$ corrections (represented by the graphs in Figs.~\ref{fig:boxes}
(d)--(f) and~\ref{fig:pentagons}) are concerned, in the case of $d$,
$s$ and $b$ final-state quarks, only the direct diagrams
(Fig.~\ref{fig:boxes} (d)--(e) and Fig.~\ref{fig:pentagons} (a)) are
present due to charge conservation, while, for $u$ and $c$ quarks, only
crossed diagrams are present, if the sum over initial- and final-state
helicities is taken. In the case of $ZZ$ exchange, all the graphs survive,
giving rise to a cancellation at the leading-log level between direct
and crossed diagrams, which does not occurr for $WW$ exchange. Hence,
the big negative correction is due to the presence of the
$WW$ non-factorisable graphs, which develop the aforementioned 
large Sudakov double logarithms in
the high energy regime.

We then show the impact of the EW corrections on some
differential distributions of phenomenological interest.
We organize the plots by showing the tree-level contributions and the higher order
corrections in three different 
contributions: the purely weak-interaction contribution 
(labelled ``weak ${\cal O}(\alpha)$"), purely weak plus QED corrections,
which are dominated by the above-mentioned ISR (labelled ``exact ${\cal
O}(\alpha)$"), and the weak plus electromagnetic 
correction in which the leading logarithms have been summed (labelled ``exact
${\cal O}(\alpha)$ + h.o.LL"). The figures show in the upper panel
the absolute distributions and in the lower panel the relative differences with
respect to the tree-level rates.
\FIGURE[t]{
\includegraphics[width=12.5cm]{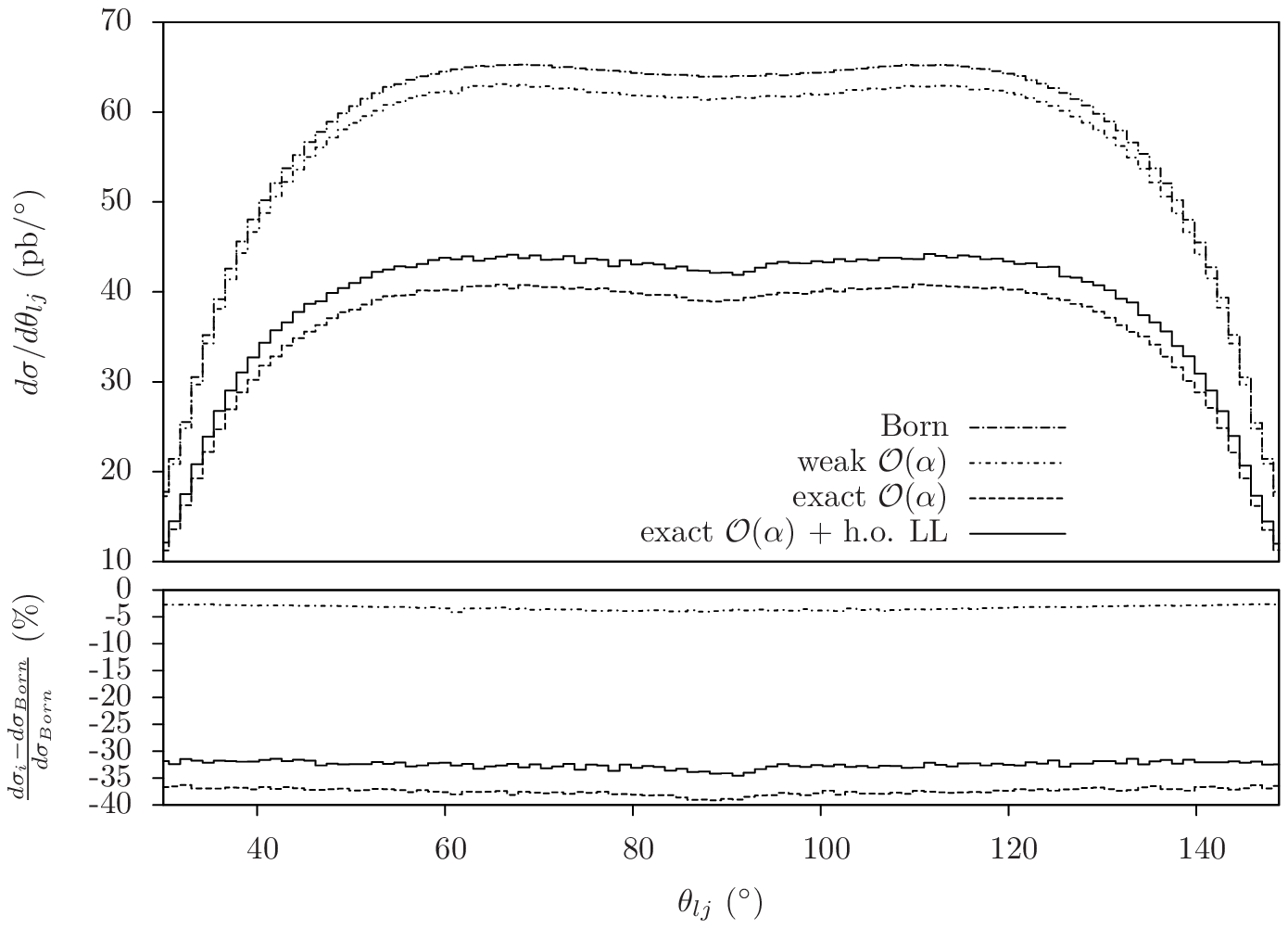}
\caption{Leading jet angle distribution at the $Z$ peak.}
\label{ajl-peak}}
\FIGURE[t]{
\includegraphics[width=12.5cm]{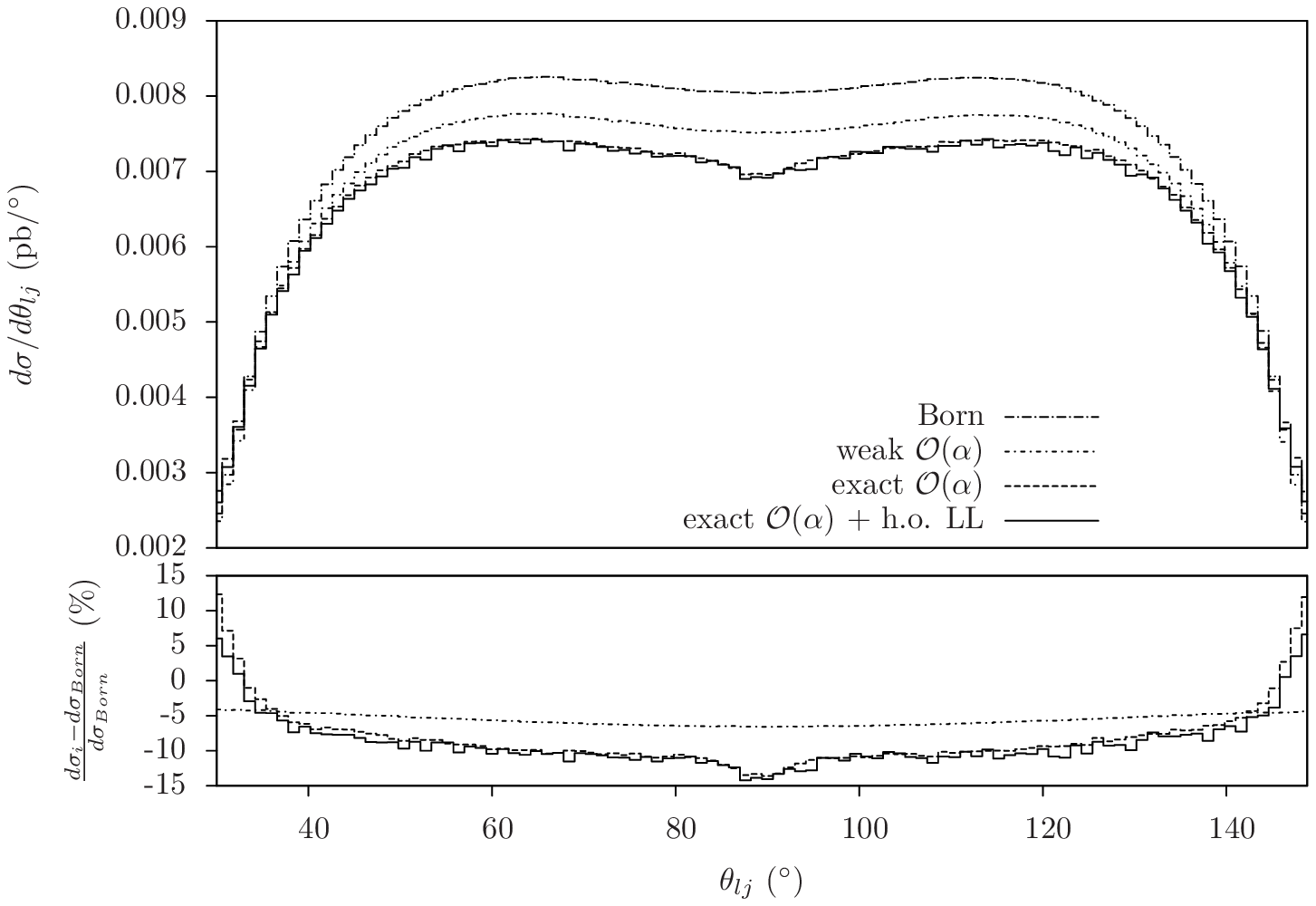}
\caption{Leading jet angle distribution at 350 GeV.}
\label{ajl-350}}
\FIGURE[t]{
\includegraphics[width=12.5cm]{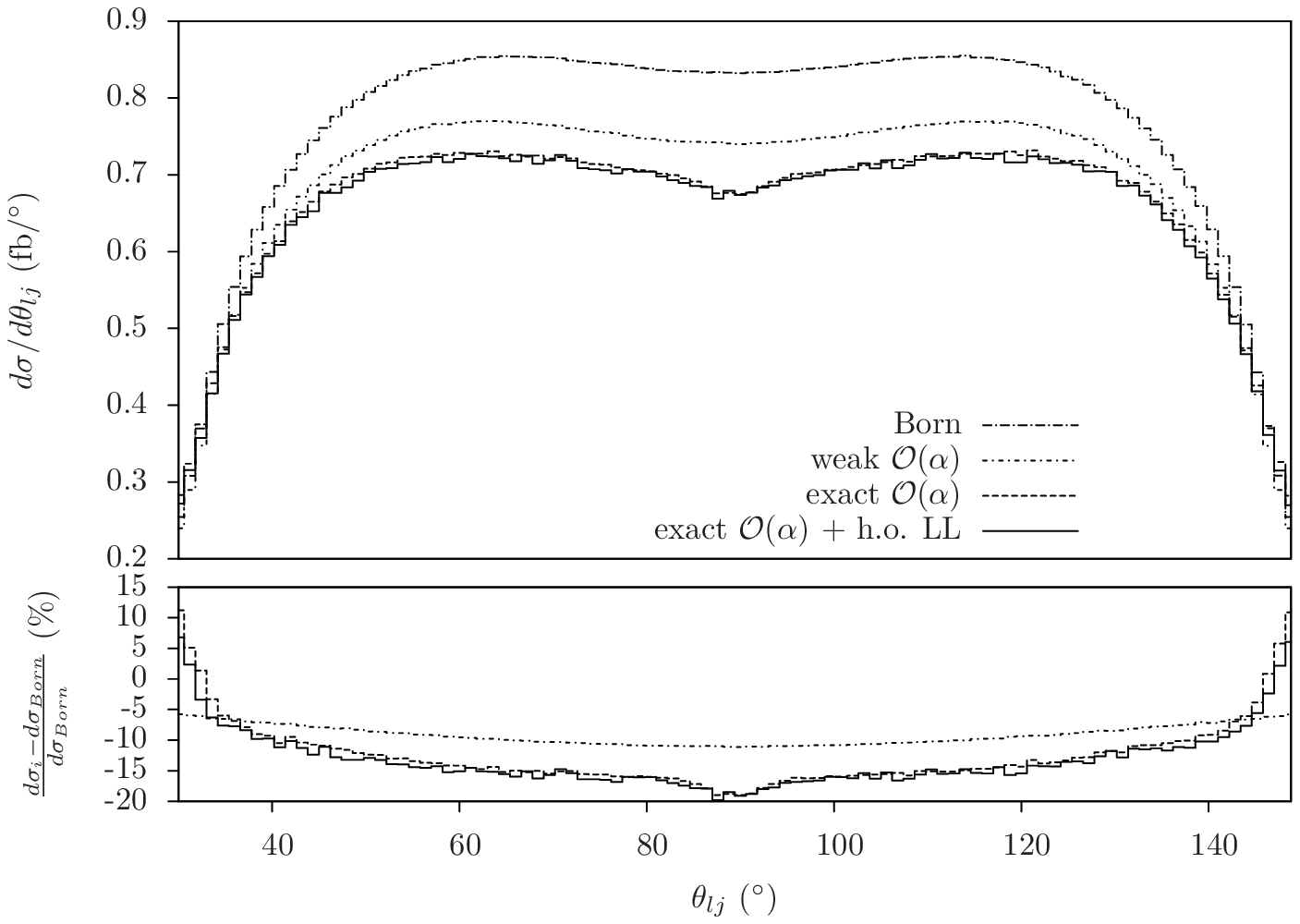}
\caption{Leading jet angle distribution at 1 TeV.}
\label{ajl-1000}}
In Fig.~\ref{ajl-peak}, the azimuthal angle of the leading
jet\footnote{The leading jet is defined as the most energetic one.}
distribution is shown at the $Z$ peak. Here,
the largest contribution to the total correction ($-30$\% or so) comes
from QED ISR because of the radiative return phenomenon. The purely weak
corrections are negative and at the level of $3-4\%$. The higher-order QED
radiation tends to compensate the order $\alpha_{\rm{EM}}$ effect, since it 
enhances
the cross section by $\sim 5$\%. By raising the CM energy,
the relative weight of the weak corrections becomes more important
(see Figs.~\ref{ajl-350} and~\ref{ajl-1000} for the leading jet angle
distribution at $\sqrt{s}=350$~GeV
and $1$~TeV, respectively) and the effect of higher-order QED
corrections becomes negligible. It is worth noticing that, far from the
$Z$ peak, the cut $M_{3j}>0.75\times\sqrt{s}$ is more effective in
reducing the radiative return phenomenon, reducing in turn the
relative effect of ISR.

\FIGURE[t]{
\includegraphics[width=12.5cm]{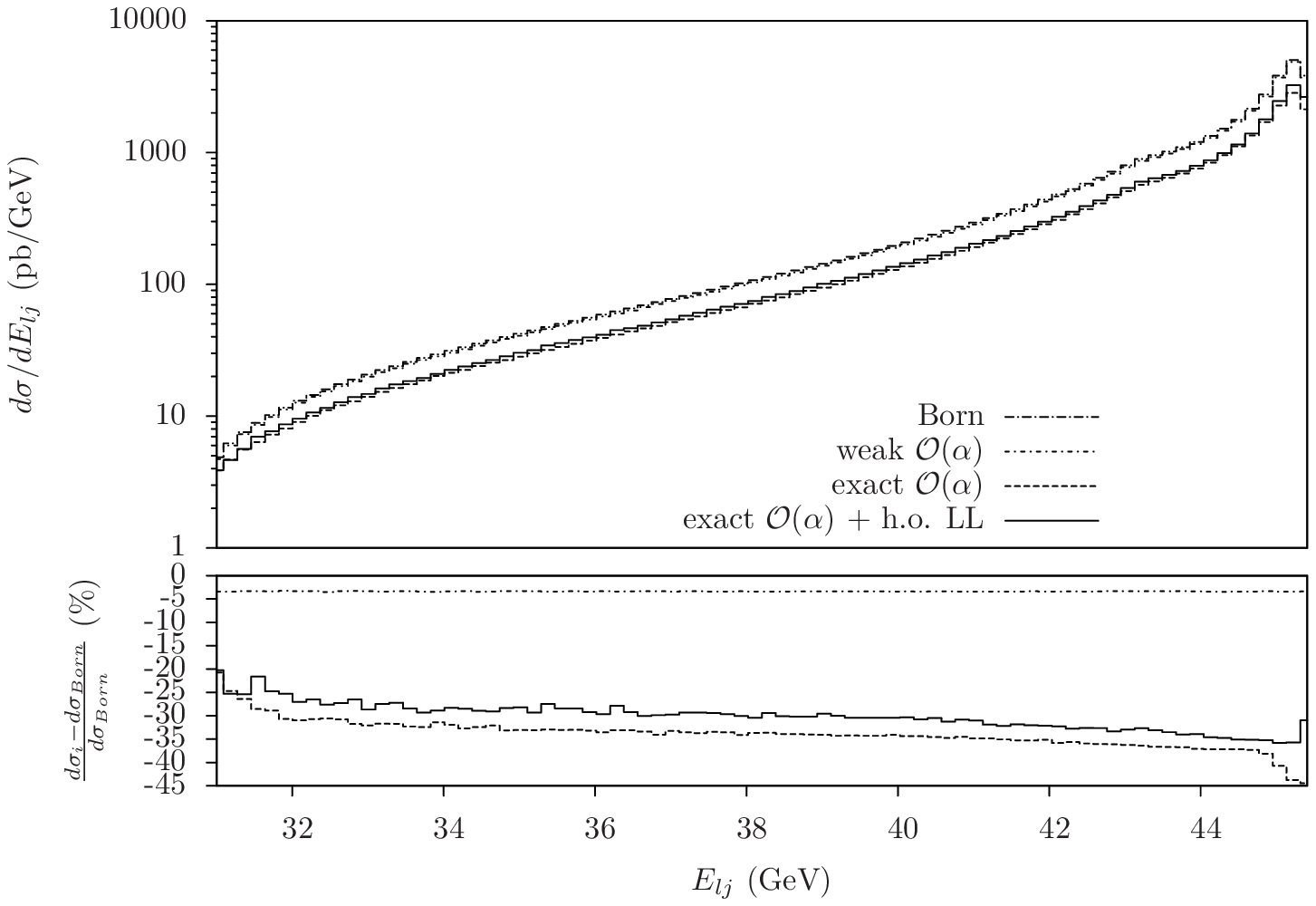}
\caption{Leading jet energy distribution at the $Z$ peak.}
\label{ejl-peak}}
\FIGURE[t]{
\includegraphics[width=12.5cm]{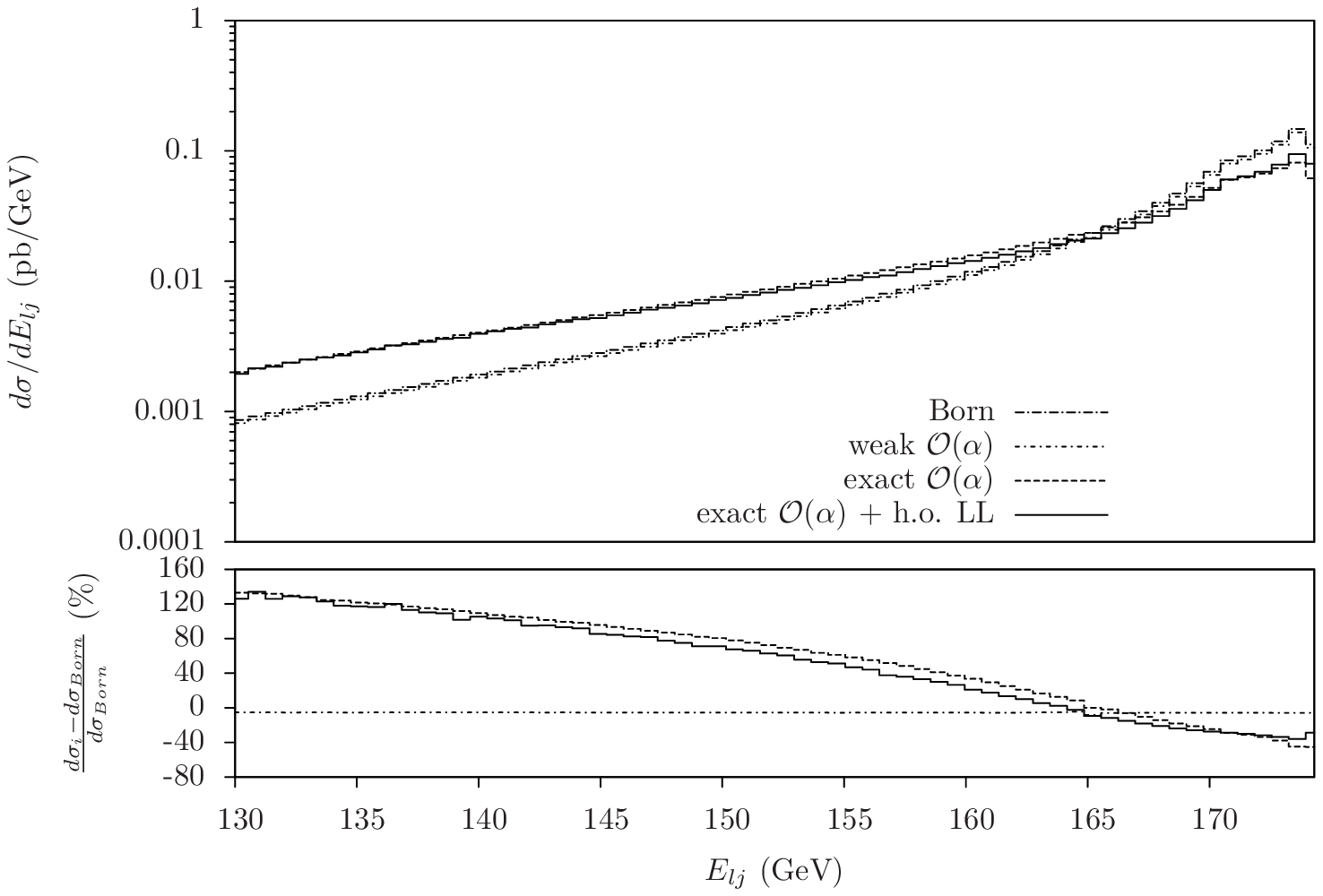}
\caption{Leading jet energy distribution at 350 GeV.}
\label{ejl-350}}
\FIGURE[t]{
\includegraphics[width=12.5cm]{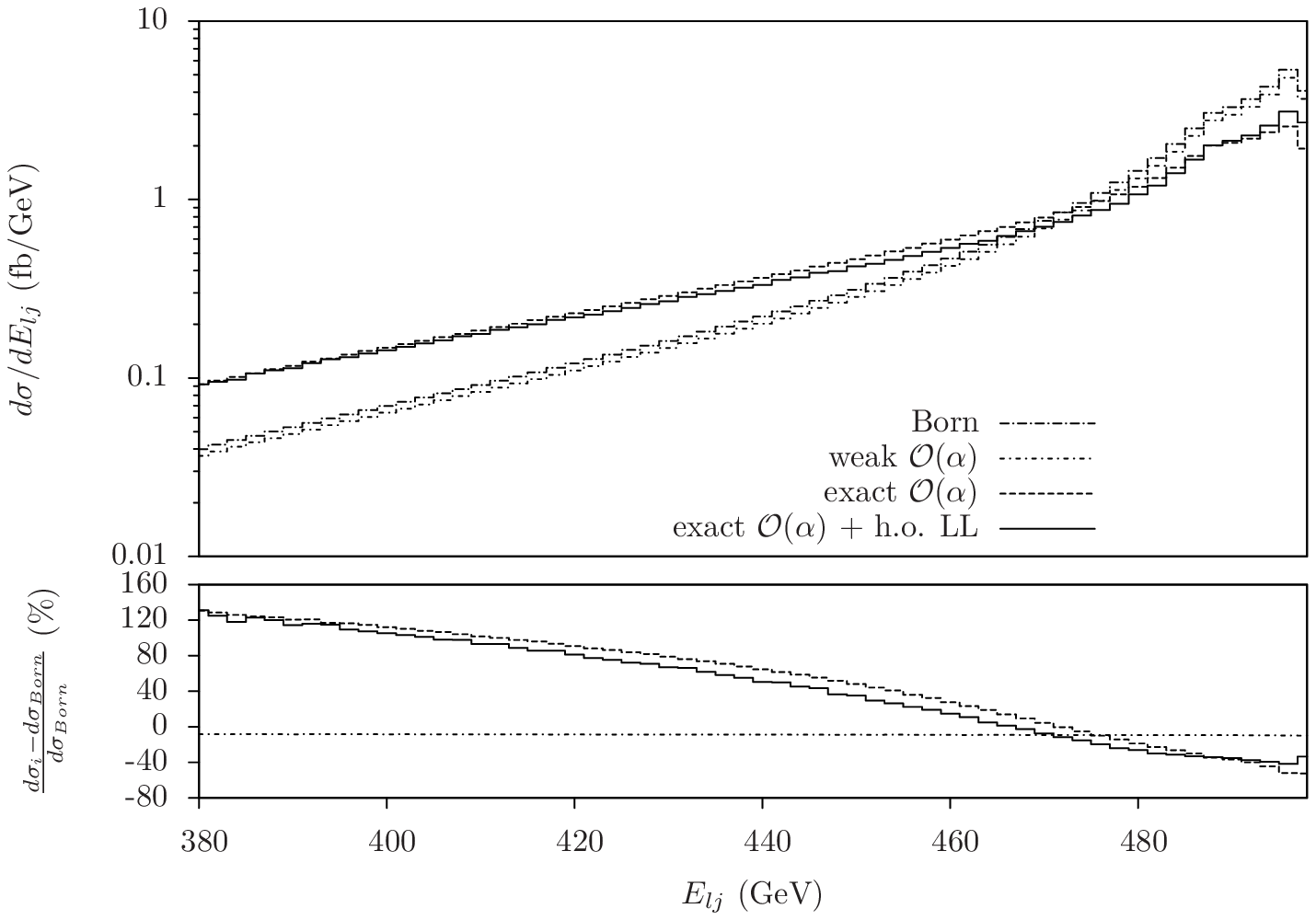}
\caption{Leading jet energy distribution at 1 TeV.}
\label{ejl-1000}}
In Figs.~\ref{ejl-peak},~\ref{ejl-350} and~\ref{ejl-1000} the
leading jet energy is presented at $\sqrt{s}=M_Z$,
$350$~GeV and $1$~TeV. This distribution is much more sensitive to the
real radiation, as it can be expected. The corrections are very large
in the distribution tail, where however the cross section is small.

\FIGURE[t]{
\includegraphics[width=12.5cm]{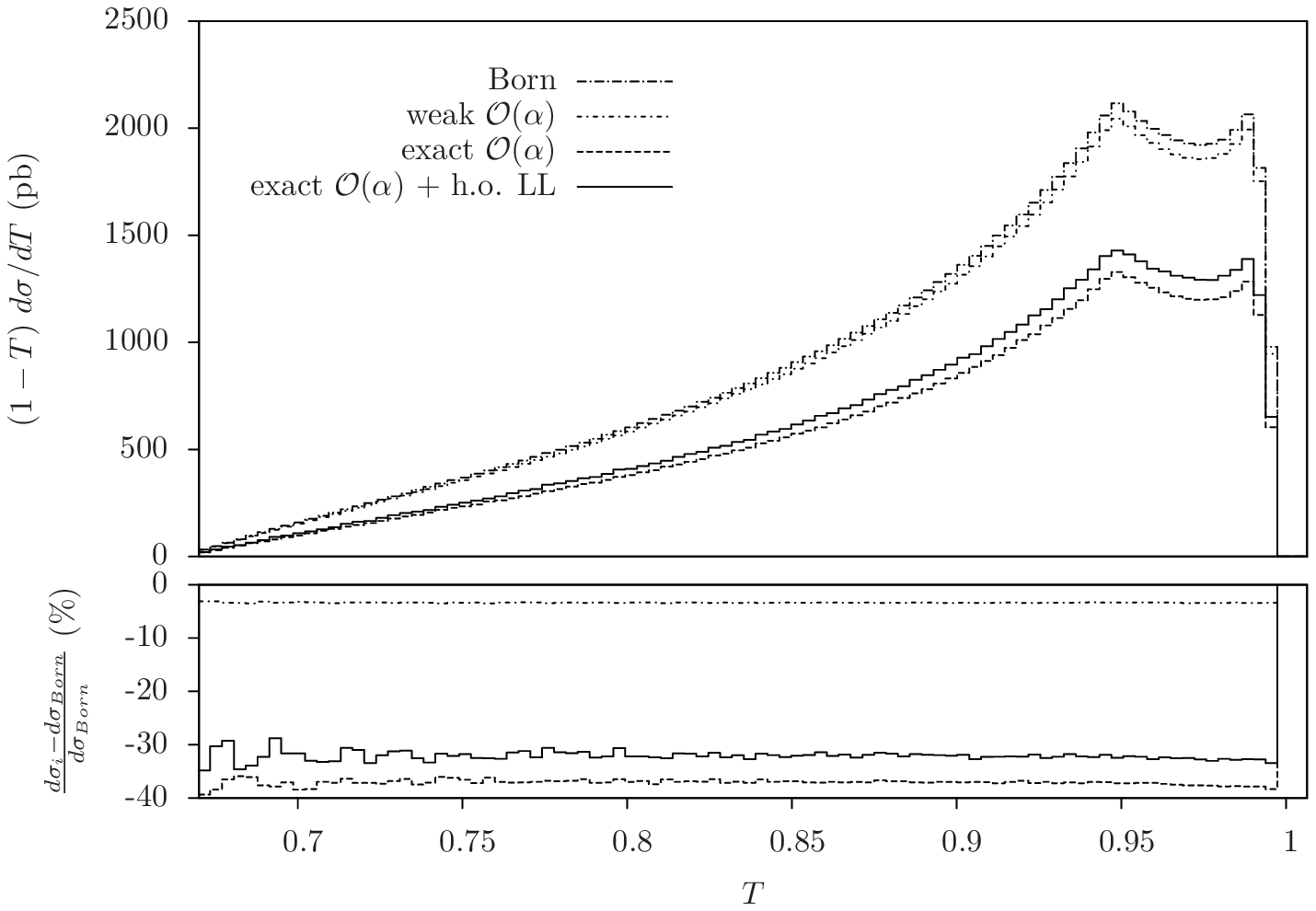}
\caption{$(1-T)\frac{d\sigma}{dT}$ distribution at the $Z$ peak.}
\label{thrust-peak}}
\FIGURE[t]{
\includegraphics[width=12.5cm]{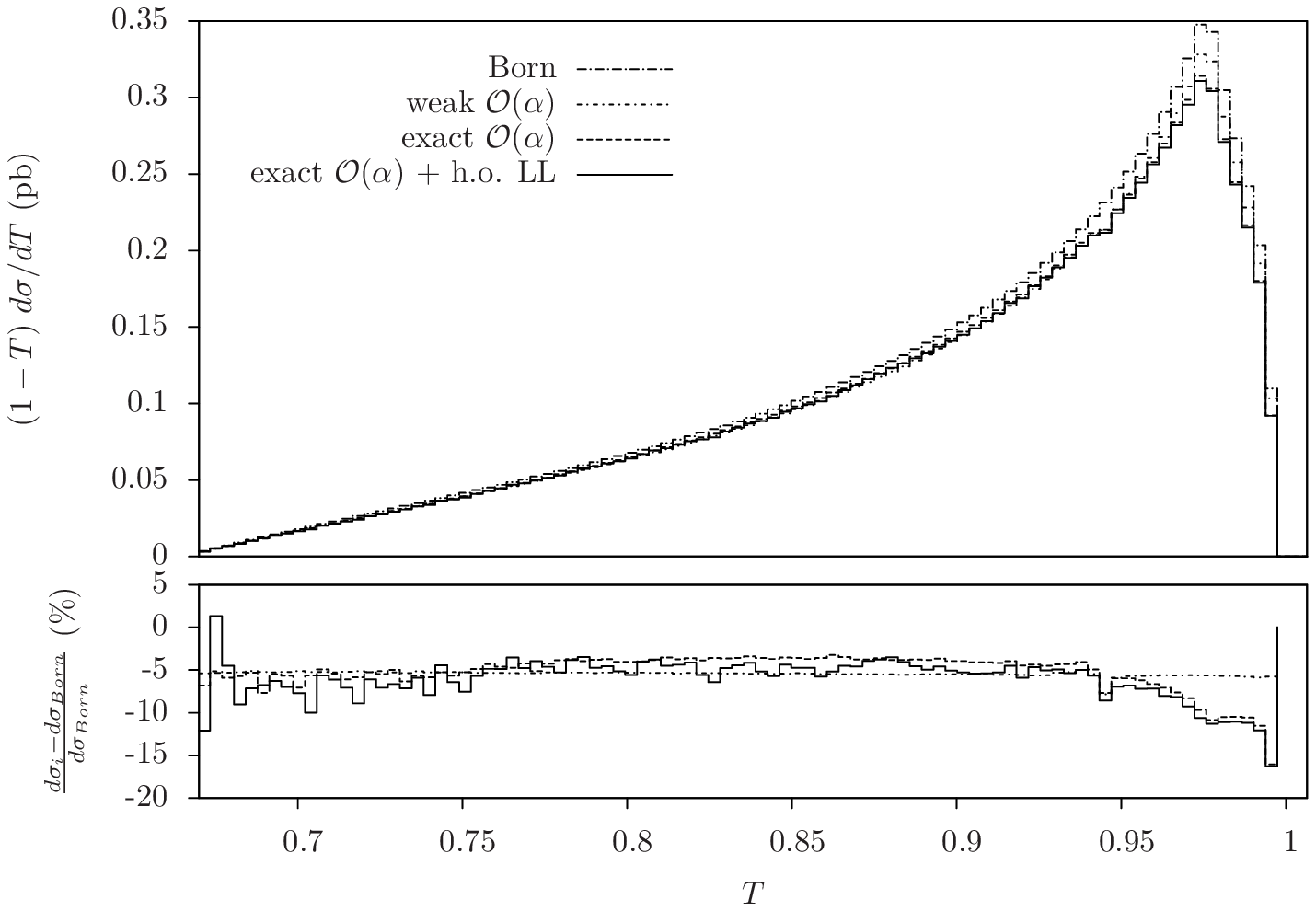}
\caption{$(1-T)\frac{d\sigma}{dT}$ distribution at 350 GeV.}
\label{thrust-350}}
\FIGURE[t]{
\includegraphics[width=12.5cm]{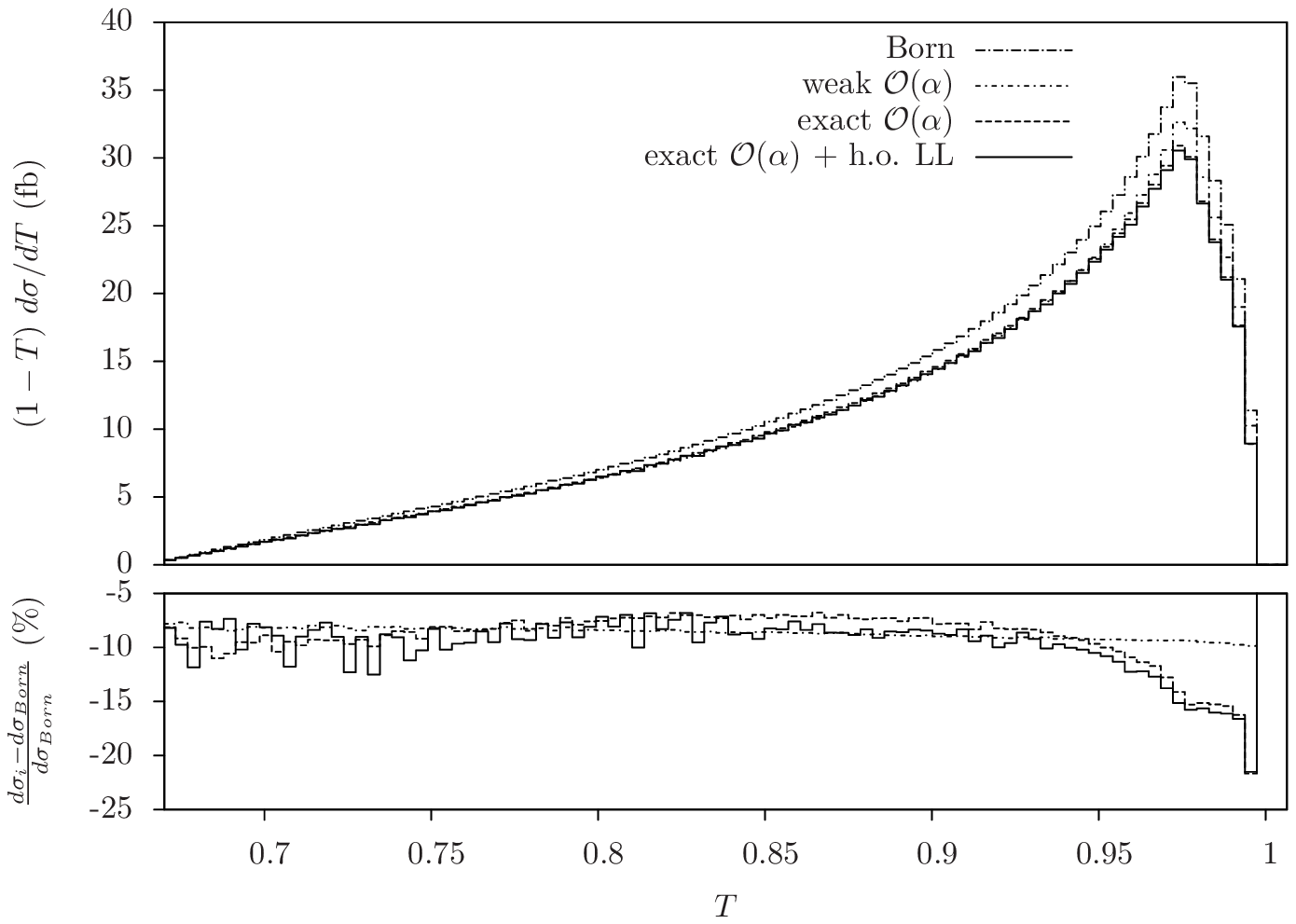}
\caption{$(1-T)\frac{d\sigma}{dT}$ distribution at 1 TeV.}
\label{thrust-1000}}
In the following some event shape variables are 
considered: the {\em thrust} $T$~\cite{thrust}, 
the {\em spherocity} $S$~\cite{spherocity}, 
the {\em C-}parameter~\cite{cparam} and the {\em oblateness} 
$O$~\cite{oblateness}. One should notice that, owning to the fact
that -- as intimated -- we are not computing here 
${\cal O}(\alpha_{\rm S}\alpha_{\rm{EM}}^3)$
one-loop QCD corrections to $q\bar q\gamma$ final states necessary
to remove singularities associated with IR real gluon emission
from $q\bar qg\gamma$ ones, we ought to 
maintain a $y_{cut}$ between jets (we use our default value), thereby
defining 
the shape variables using the jets three-momenta, which at the same time
enables us to remove energetic and isolated `photon jets' from the final
state. (In fact, for consistency, `photon jets' are not used in the
calculation of the shape variables even when the gluon is resolved
in a $q\bar qg\gamma$ final state.)  
This is not unlike typical experimental procedures, whereby resolved photons
from either the ISR or from the final state (i.e., those not originating
from parton fragmentation) are separated from the hadronic
system by using a jet clustering algorithm\footnote{This procedure
is in fact standard at LEP2, see, e.g., Ref.~\cite{ALEPH}, where
resolved photons from both the initial and final states are
more numerous than at LEP1, where the $Z$ width naturally suppresses
the former and where the phase space is more limited for the latter.}. 
In Figs.~\ref{thrust-peak},~\ref{thrust-350} and~\ref{thrust-1000},
the distribution $(1-T)\frac{d\sigma}{dT}$ is shown. $T$ is 
defined as
\begin{equation}
T ={\mathrm {max}}_{\vec{n}}
\frac{\sum_i|\vec{p_i}\cdot\vec{n}|}{\sum_i|\vec{p_i}|},
\label{eq:thrust}
\end{equation}
where the sum runs over the reconstructed jet momenta
and $\vec{n}$ is the thrust axis. 
The $T$ distribution is one
of the key observables used for the measurement of $\alpha_{\rm S}$ in $e^+e^-$
collisions~\cite{kunsztnason}. It is worth noticing that while the
purely weak corrections give an almost constant effect on the whole
$T$ range, the presence of the real bremsstrahlung gives a non trivial
effect in the region $T>0.92$. In view of a precise measurement of
$\alpha_{\rm S}$ at a future LC, EW corrections can
play an important role.

\FIGURE[t]{
\includegraphics[width=12.5cm]{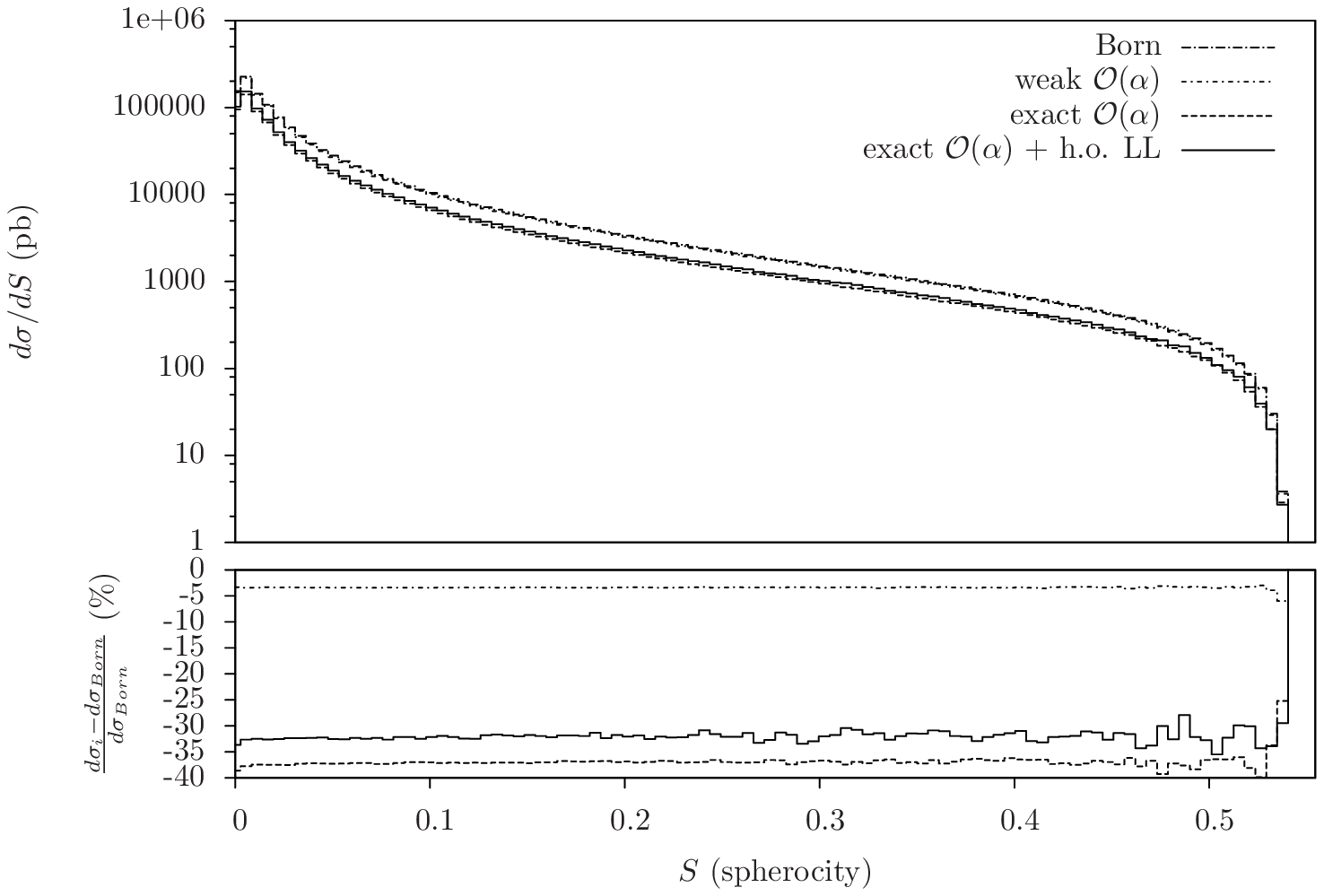}
\caption{$\frac{d\sigma}{dS}$ distribution at the $Z$ peak.}
\label{spherocity-peak}}
\FIGURE[t]{
\includegraphics[width=12.5cm]{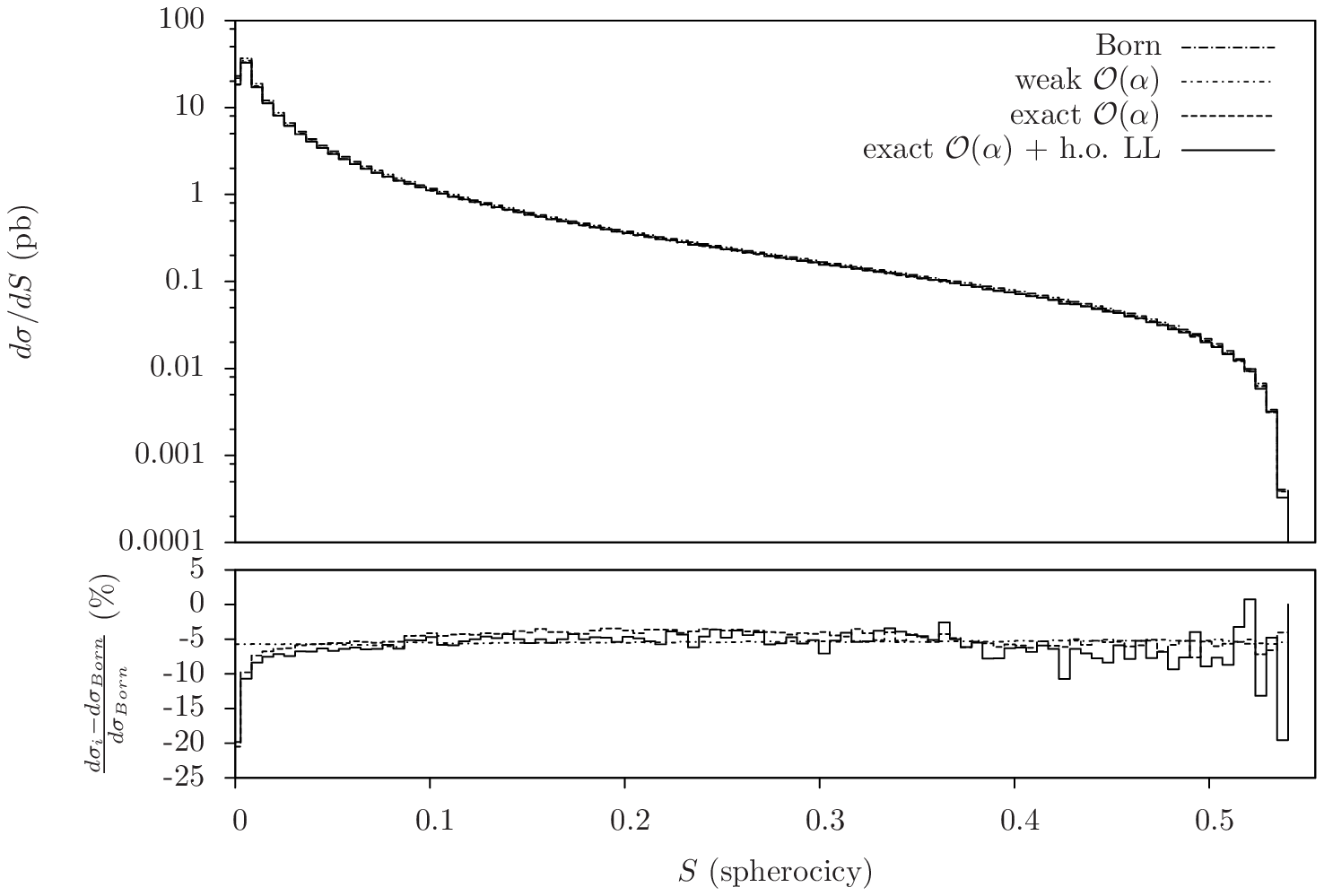}
\caption{$\frac{d\sigma}{dS}$ distribution at 350 GeV.}
\label{spherocity-350}}
\FIGURE[t]{
\includegraphics[width=12.5cm]{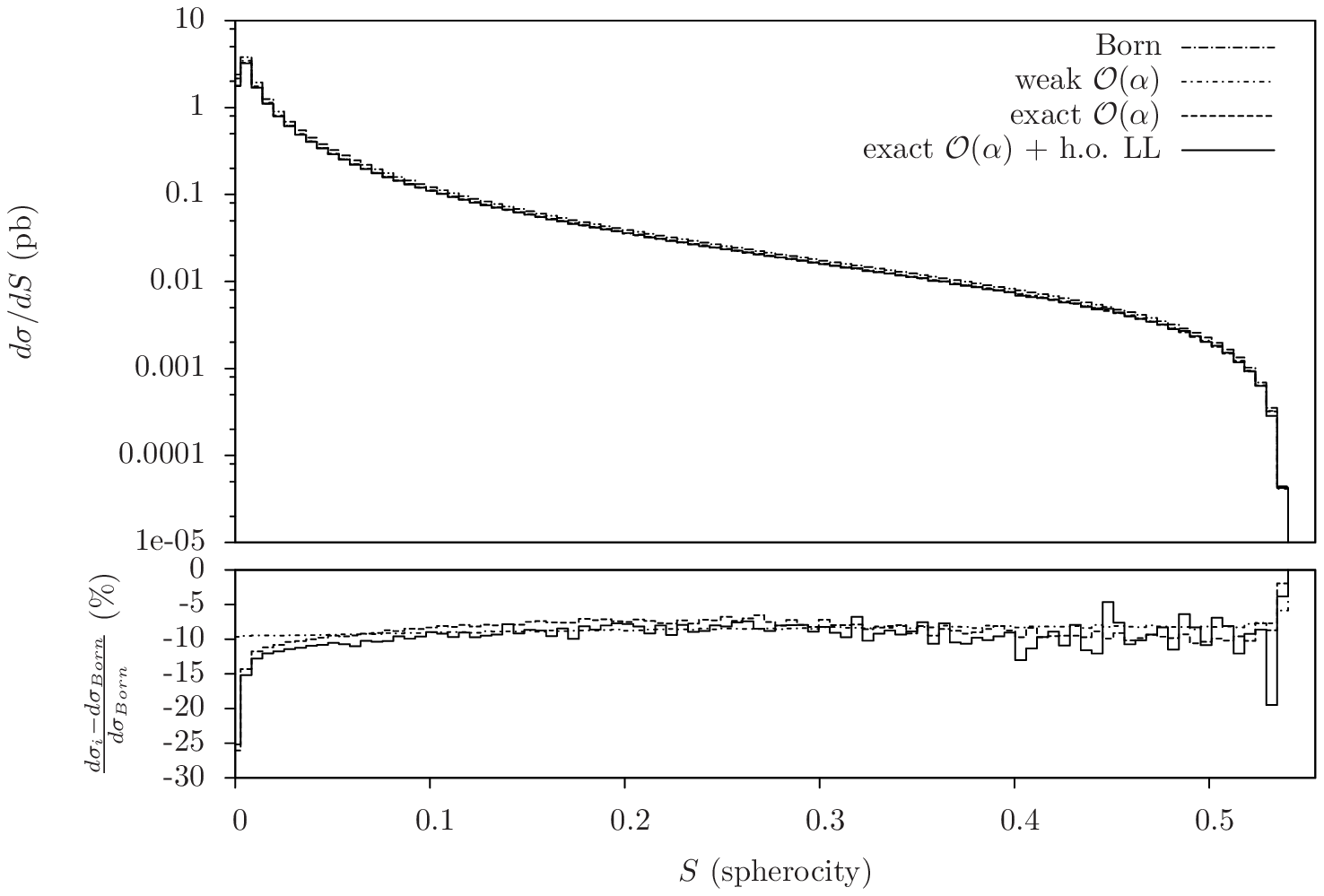}
\caption{$\frac{d\sigma}{dS}$ distribution at 1 TeV.}
\label{spherocity-1000}}
In Figs.~\ref{spherocity-peak},~\ref{spherocity-350} 
and~\ref{spherocity-1000},
the distribution $\frac{d\sigma}{dS}$ is shown. $S$ is 
defined as
\begin{equation}
S =\left( \frac {4}{\pi}\right)^2 {\mathrm {min}}_{\vec{n}}
\frac{\sum_i|\vec{p_i}\times \vec{n}|}{\sum_i|\vec{p_i}|},
\label{eq:spherocity}
\end{equation}
where the sum runs over the reconstructed jet momenta
and $\vec{n}$ is the spherocity axis. The effect of 
pure weak corrections is flat over the whole $S$ range 
increasing in size from few percent at the $Z$ peak 
to about 10\% at 1~TeV. The addition of QED radiation 
at ${\cal O}(\alpha)$ and at all orders 
does not change the picture except for the $Z$ peak, where 
the ${\cal O}(\alpha)$ contribution is about 30\% and the 
higher order corrections give an additional 10\% effect. 

\FIGURE[t]{
\includegraphics[width=12.5cm]{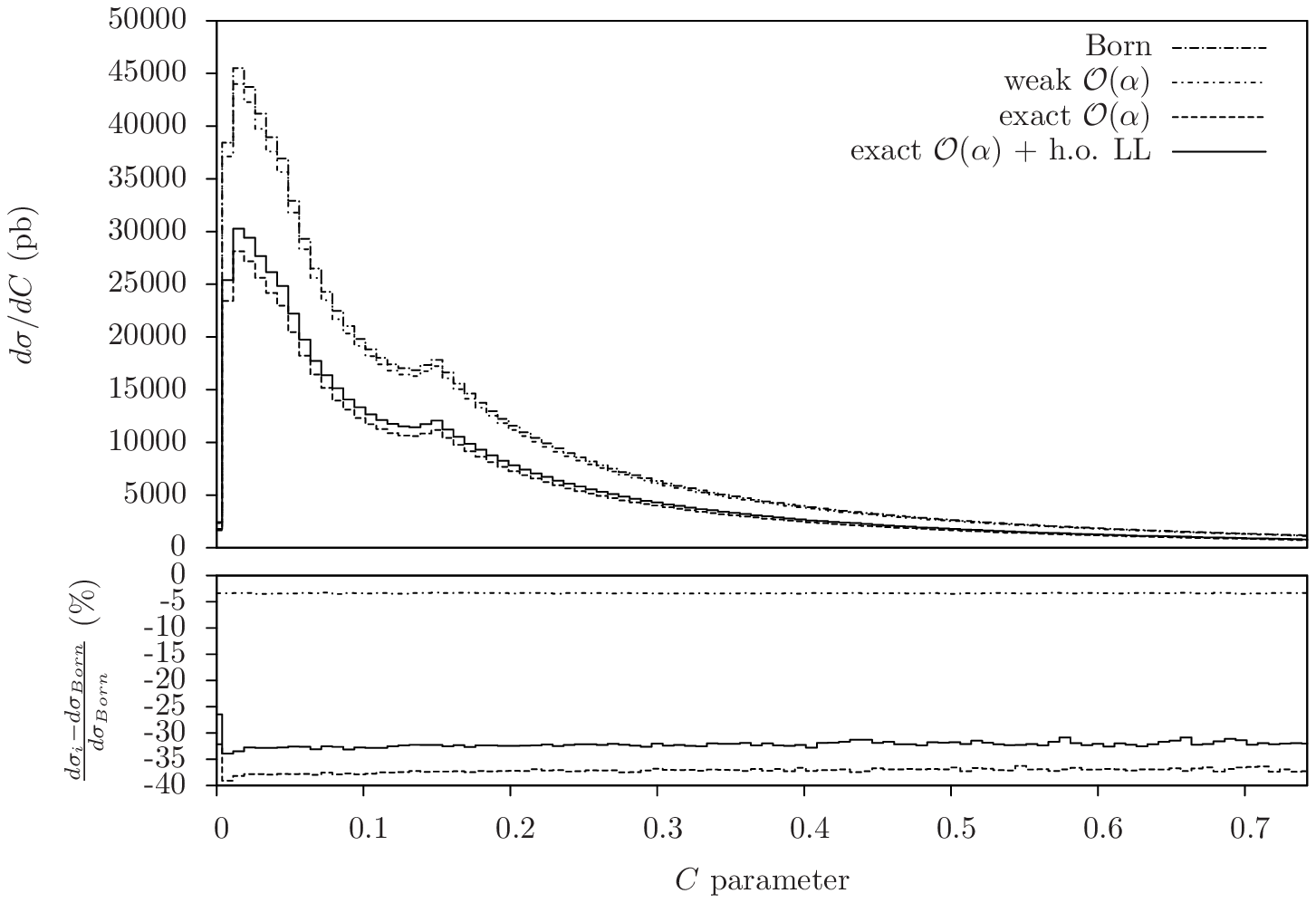}
\caption{$\frac{d\sigma}{dC}$ distribution at the $Z$ peak.}
\label{cpar-peak}}
\FIGURE[t]{
\includegraphics[width=12.5cm]{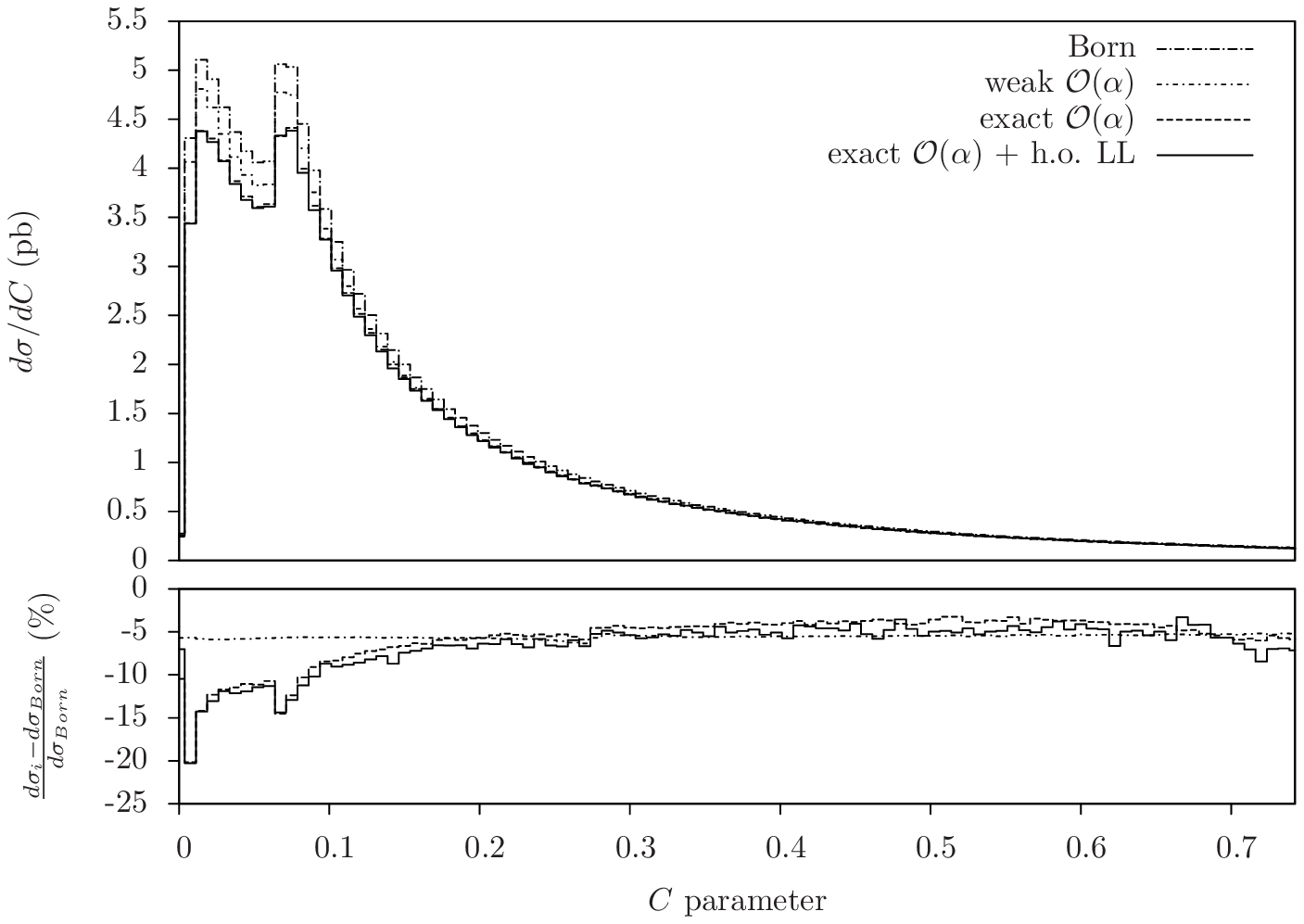}
\caption{$\frac{d\sigma}{dC}$ distribution at 350 GeV.}
\label{cpar-350}}
\FIGURE[t]{
\includegraphics[width=12.5cm]{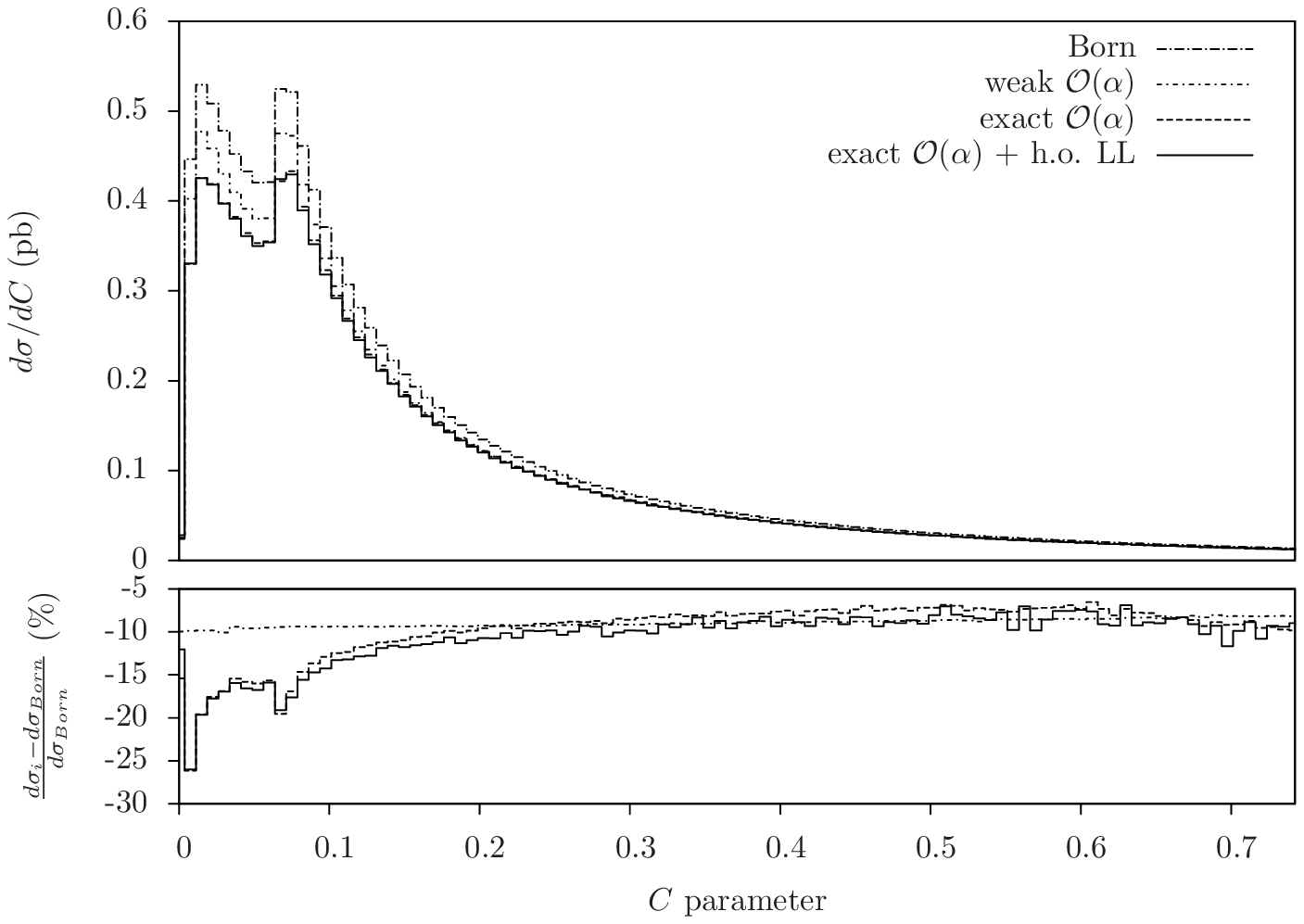}
\caption{$\frac{d\sigma}{dC}$ distribution at 1 TeV.}
\label{cpar-1000}}
In Figs.~\ref{cpar-peak},~\ref{cpar-350} 
and~\ref{cpar-1000},
the distribution $\frac{d\sigma}{dC}$ is shown. $C$ is 
defined as
\begin{equation}
C =\frac {3}{2}
\frac{\sum_{i,j}[|\vec{p_i}||\vec{p_j}|-(\vec{p_i} \cdot \vec{p_j})^2/|\vec{p_i}||\vec{p_j}|]}{\left( 
\sum_i|\vec{p_i}|\right)^2},
\label{eq:cpar}
\end{equation}
where the sum runs over the reconstructed jet momenta. 
The effect of pure weak corrections is similar in magnitude 
and shape to the case of spherocity and thrust. The addition 
of QED radiation is very large only at the $Z$ peak ($\sim 30$\%) while 
for higher energies it introduces a non trivial shape in the 
region $C \lsim 0.15$.

\FIGURE[t]{
\includegraphics[width=12.5cm]{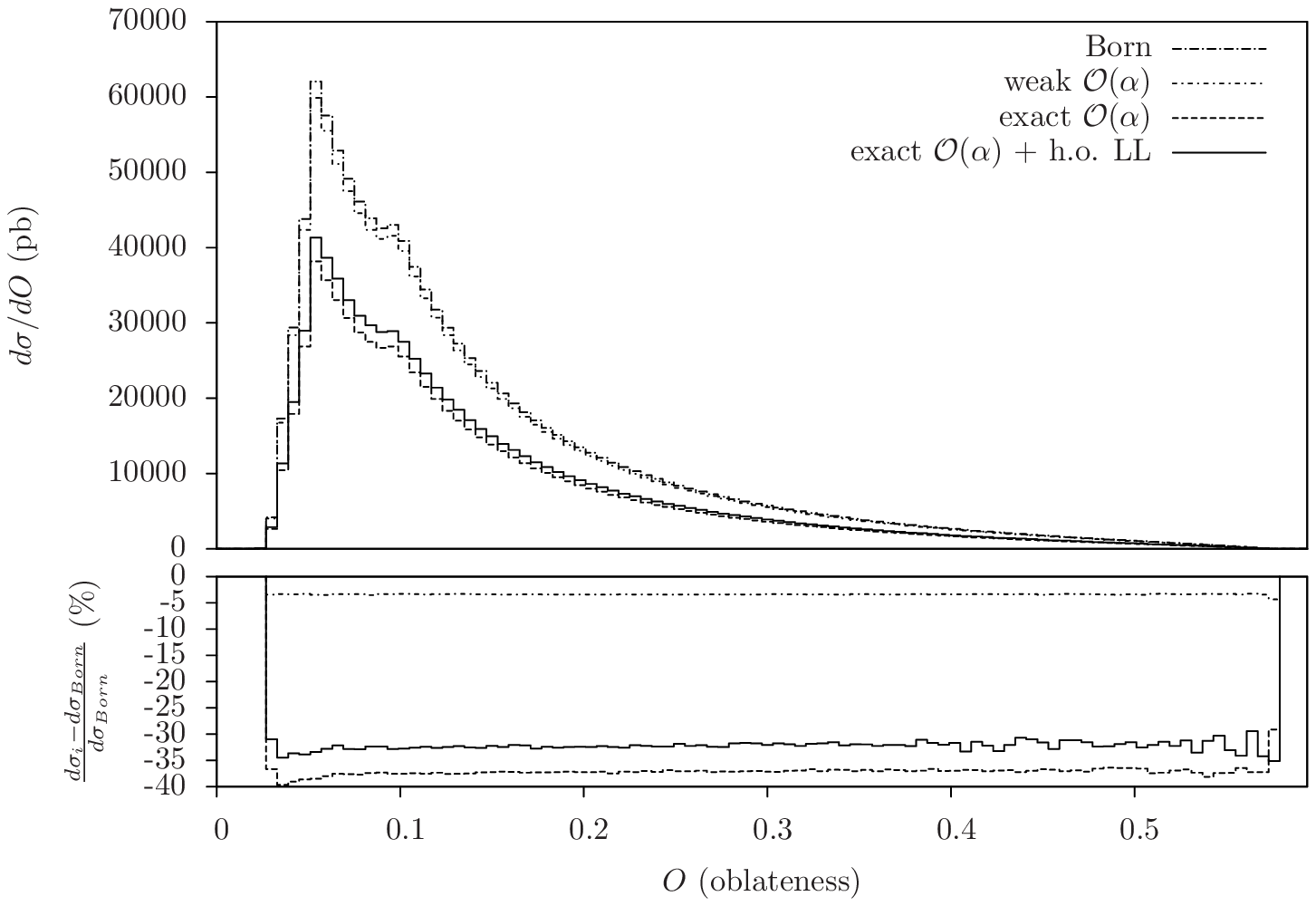}
\caption{$\frac{d\sigma}{dO}$ distribution at the $Z$ peak.}
\label{obl-peak}}
\FIGURE[t]{
\includegraphics[width=12.5cm]{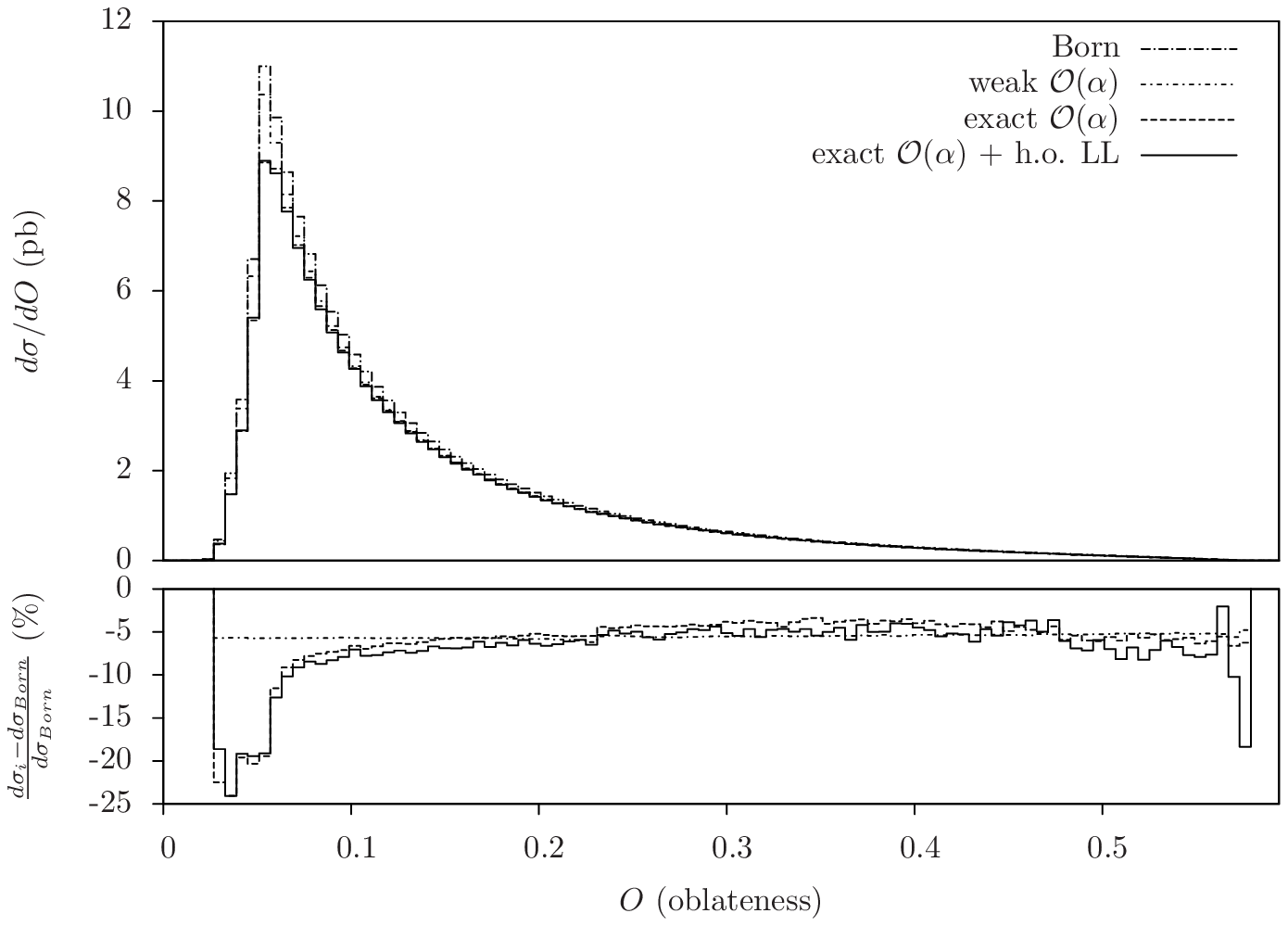}
\caption{$\frac{d\sigma}{dO}$ distribution at 350 GeV.}
\label{obl-350}}
\FIGURE[t]{
\includegraphics[width=12.5cm]{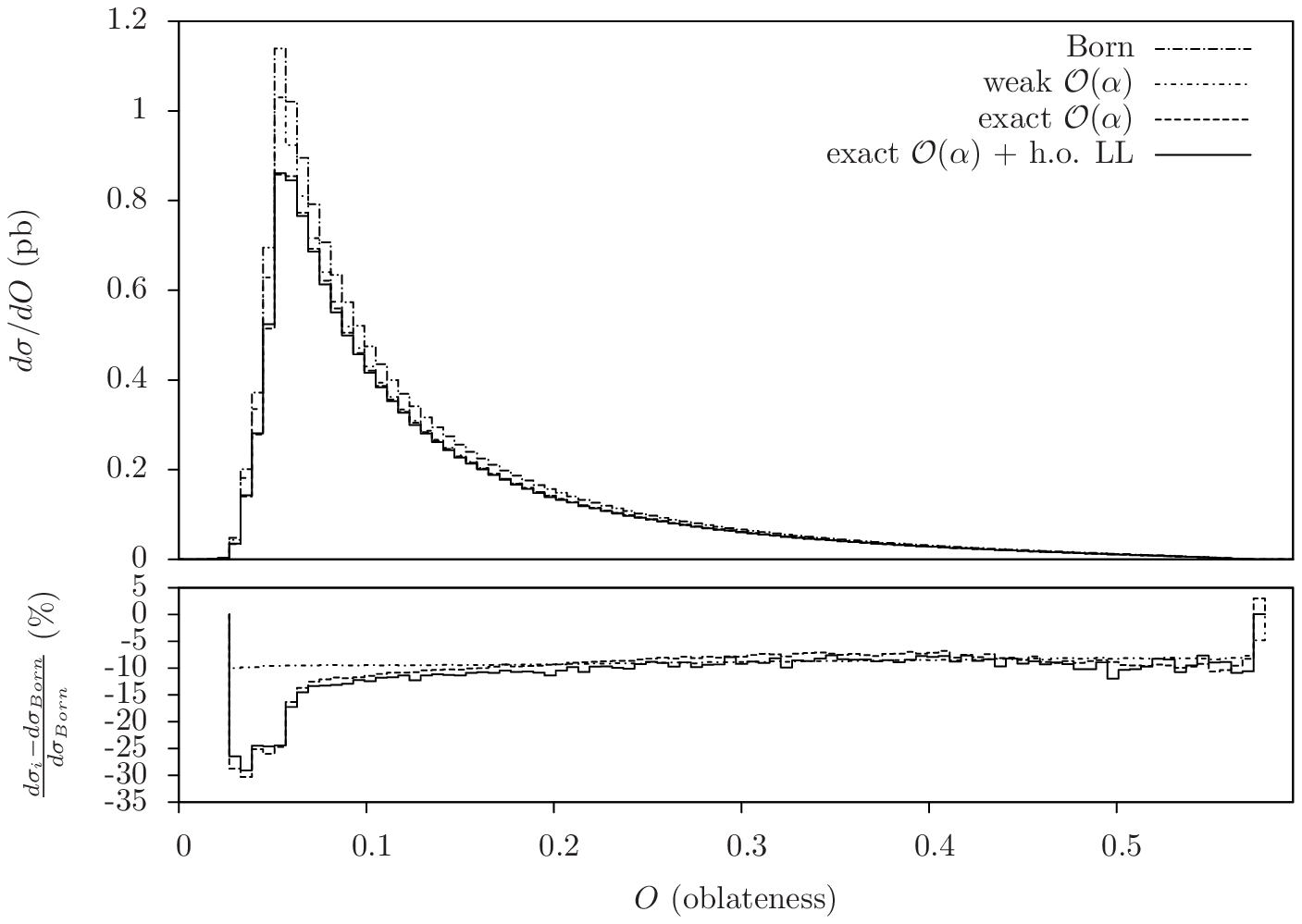}
\caption{$\frac{d\sigma}{dO}$ distribution at 1 TeV.}
\label{obl-1000}}
In Figs.~\ref{obl-peak},~\ref{obl-350} 
and~\ref{obl-1000},
the distribution $\frac{d\sigma}{dO}$ is shown. $O$ is 
defined as
\begin{eqnarray}
O &=& F_{\rm major} - F_{\rm minor} \, , \nonumber \\
F_{\rm major} &=& 
\frac{\sum_{i}|\vec{p_i} \cdot \vec{n}_{\rm major}|}{\sum_i|\vec{p_i}|}\, , 
\nonumber \\
F_{\rm minor} &=& 
\frac{\sum_{i}|\vec{p_i} \cdot \vec{n}_{\rm minor}|}{\sum_i|\vec{p_i}|}\, , 
\label{eq:obl}
\end{eqnarray}
where $\vec{n}_{\rm major}$ is an axis which lyes in the plane perpendicular 
to the thrust axis and the magnitude of the momenta projected into 
this plane is maximal along this axis; $\vec{n}_{\rm minor}$ is orthogonal 
to the thrust axis and $\vec{n}_{\rm major}$. The numerical impact 
of the various contributions is similar to the case of the 
{\em C-}parameter. The region where QED real radiation introduces 
non trivial effects is $C \lsim$~0.1.

\FIGURE[t]{
\includegraphics[width=12.5cm]{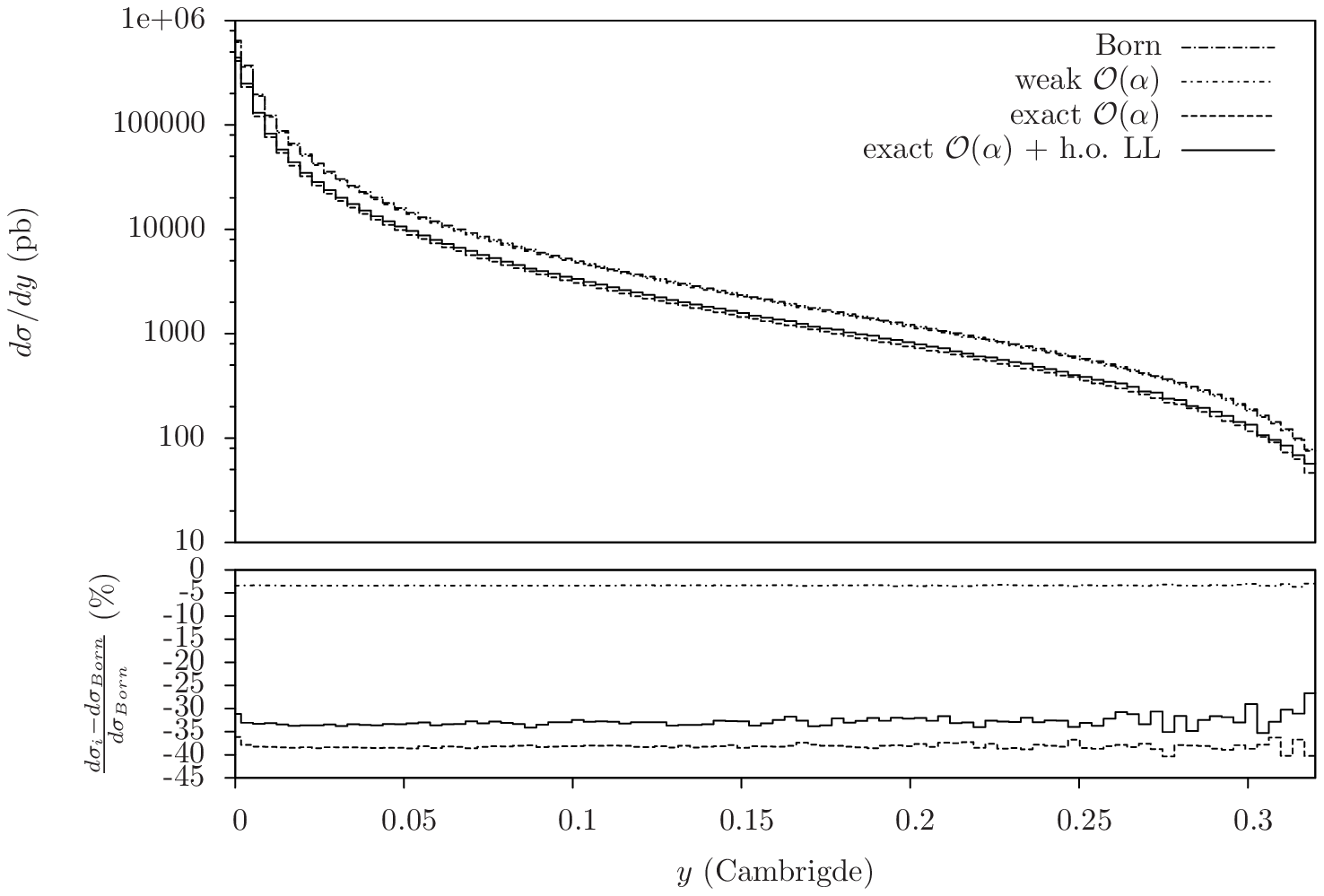}
\caption{Cambridge $y$ distribution at the $Z$ peak.}
\label{ycam-peak}}
\FIGURE[t]{
\includegraphics[width=12.5cm]{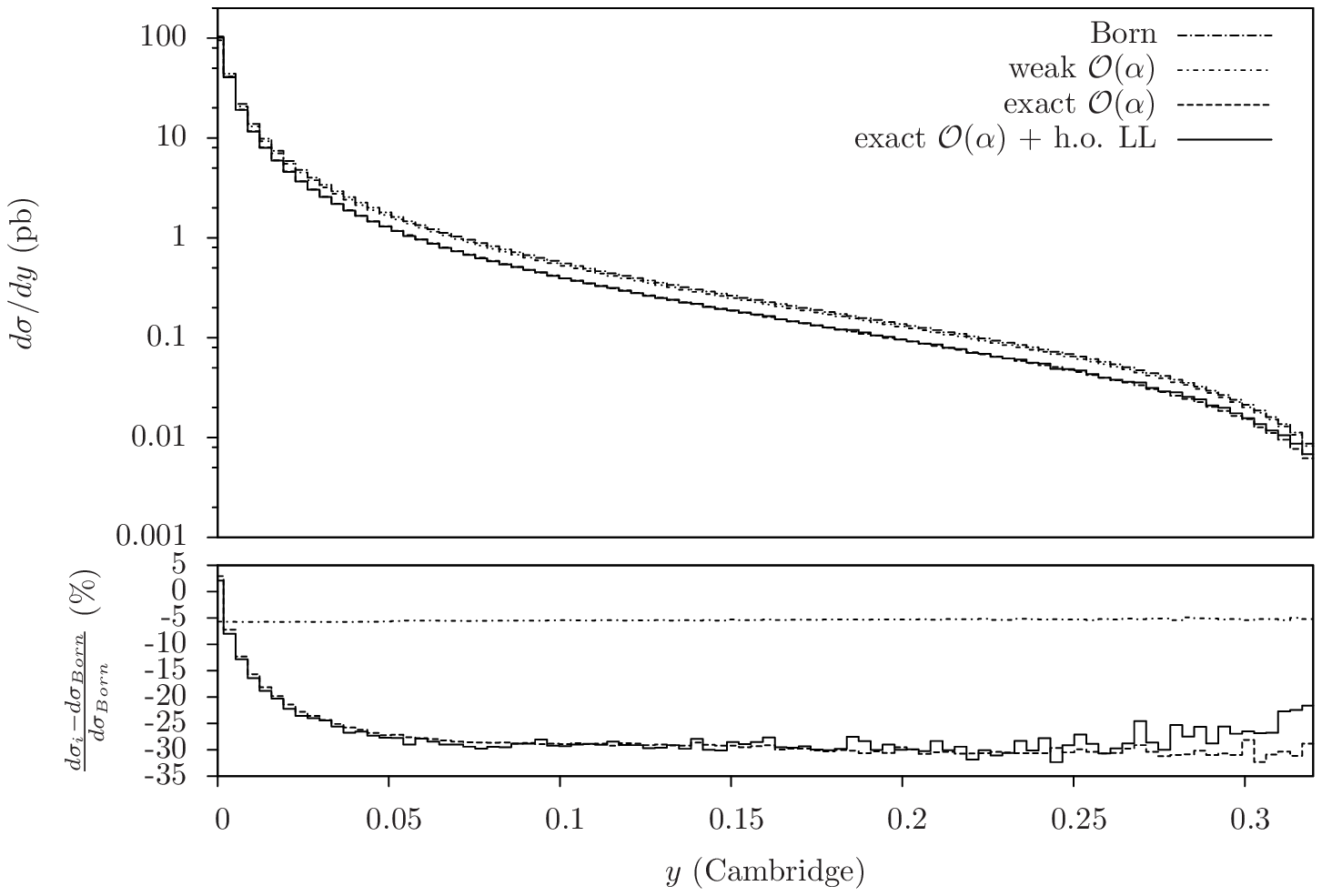}
\caption{Cambridge $y$ distribution at 350 GeV.}
\label{ycam-350}}
\FIGURE[t]{
\includegraphics[width=12.5cm]{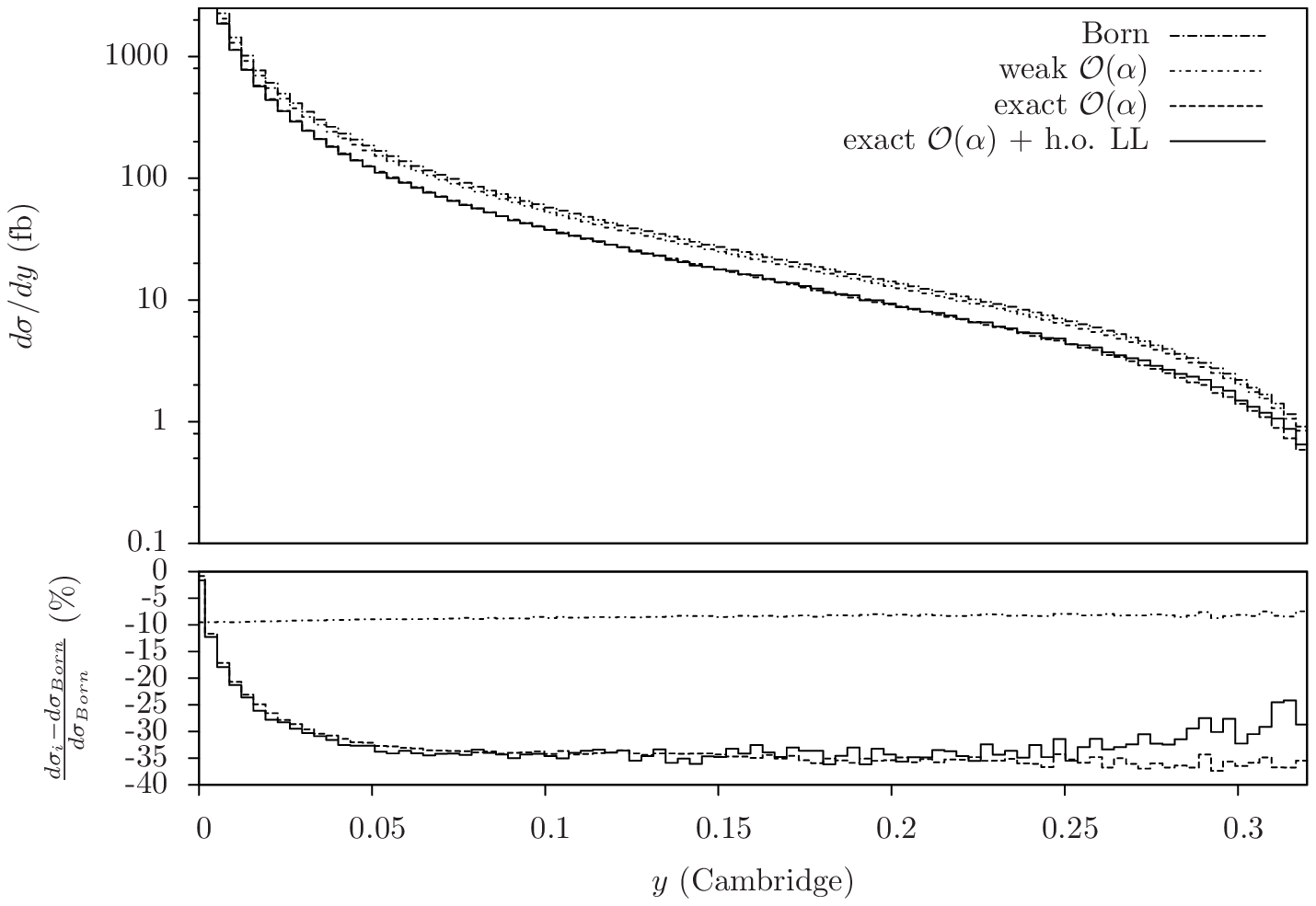}
\caption{Cambridge $y$ distribution at 1 TeV.}
\label{ycam-1000}}
In Figs.~\ref{ycam-peak}, \ref{ycam-350} and~\ref{ycam-1000}
the $y$ distribution is shown. For each event, the observable 
$y$ is defined as the minimum (Cambridge) $y_{ij}$ such
as $y_{ij}> y_{cut} = 0.001$. Also on this distribution the weak
effects are quite constant, while real radiation distorts its shape.
\FIGURE[t]{
\includegraphics[width=12.5cm]{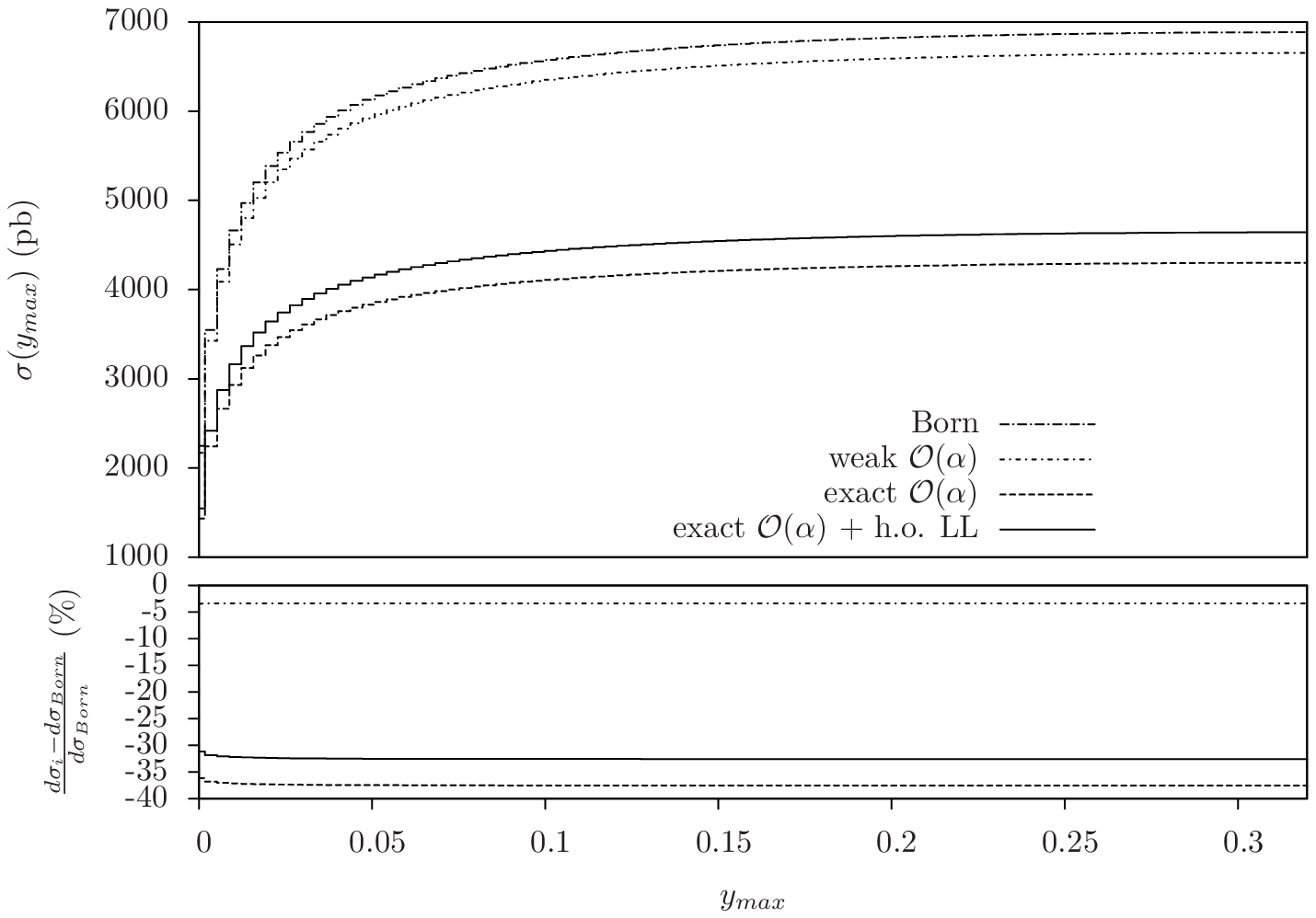}
\caption{Integrated cross section as a function of the maximum $y$ and
  relative corrections at the $Z$ peak in the Cambridge scheme.}
\label{ycamsum-peak}}
\FIGURE[t]{
\includegraphics[width=12.5cm]{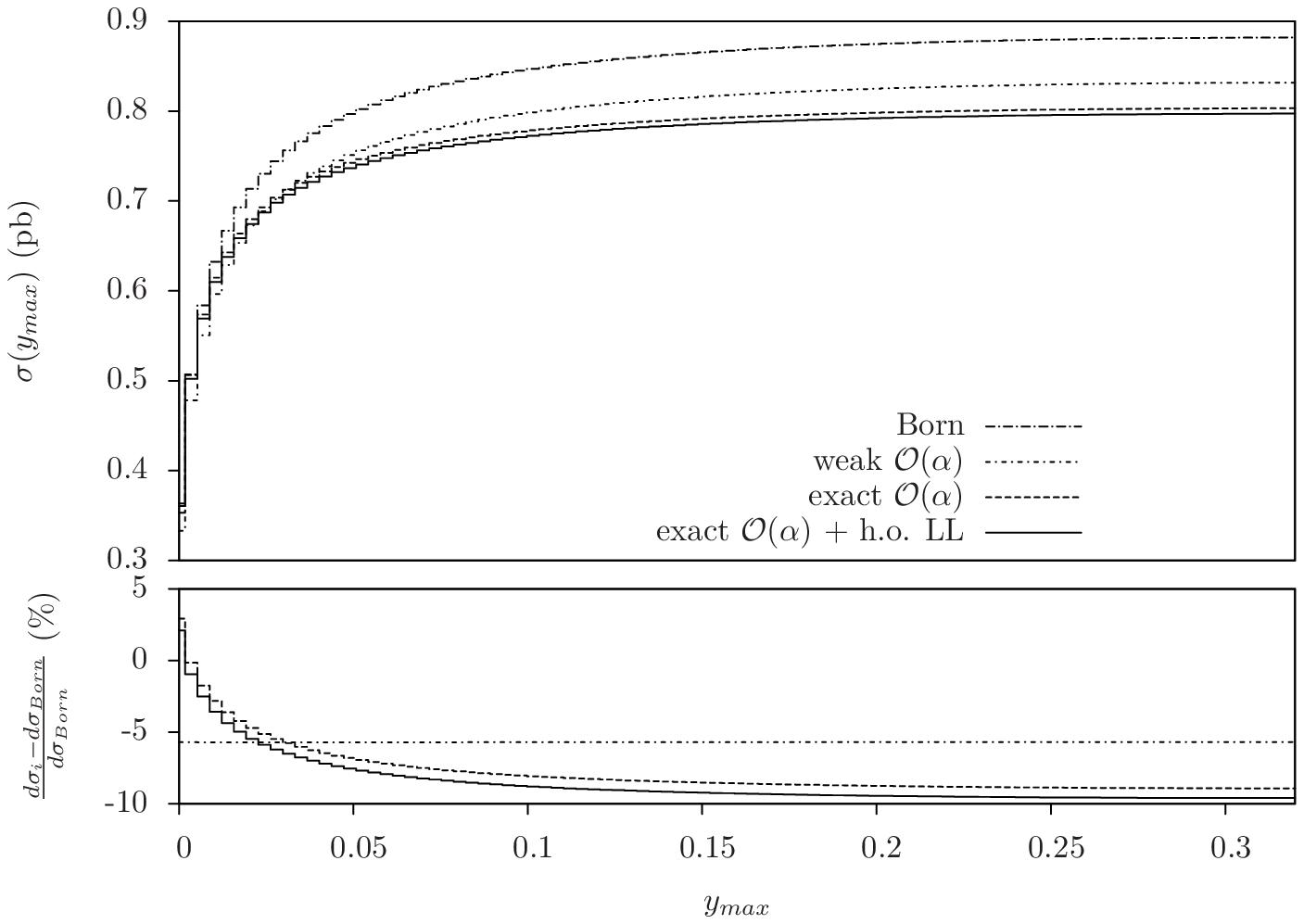}
\caption{Integrated cross section as a function of the maximum $y$ and
  relative corrections at 350 GeV in the Cambridge scheme.}
\label{ycamsum-350}}
\FIGURE[t]{
\includegraphics[width=12.5cm]{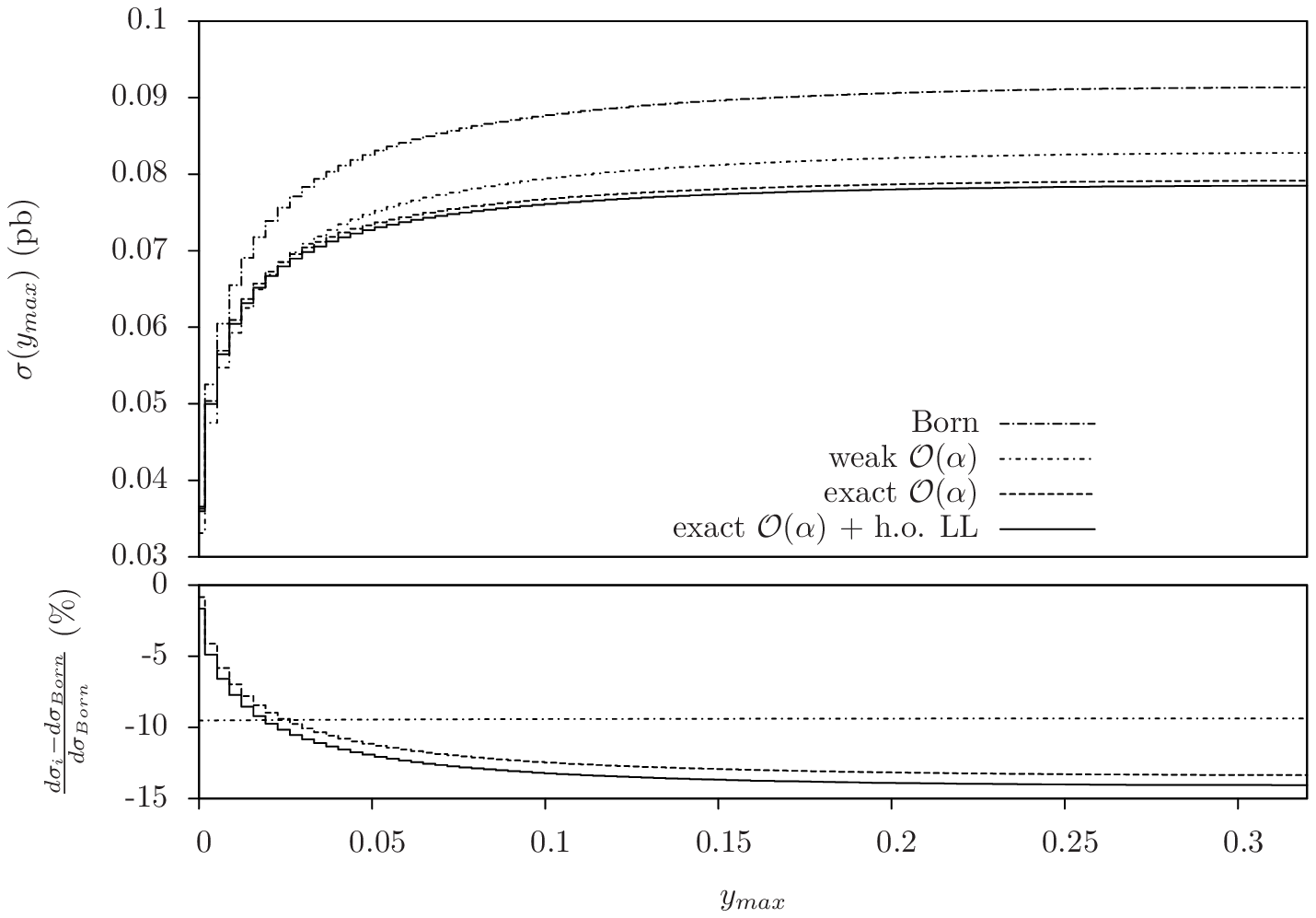}
\caption{Integrated cross section as a function of the maximum $y$ and
  relative corrections at 1 TeV  in the Cambridge scheme.}
\label{ycamsum-1000}}
Finally, Figs.~\ref{ycamsum-peak}, \ref{ycamsum-350}
and~\ref{ycamsum-1000} represent the cross sections integrated over $y$ in the
range $y_{cut}< y < y_{max}$, as a function of $y_{max}$.
Corrections can be very large in both distributions generally at any energy,
reaching the several tens of percent. 

\section{Conclusions}
\label{Sec:Conclusions}
In summary, we have shown the phenomenological relevance that the calculation
up to
${\cal O}(\alpha_{\rm S}\alpha_{\rm{EW}}^3)$
can have in the study of (unflavoured) three-jet samples
in $e^+e^-$ annihilation, for all energies ranging from $\sqrt s=M_Z$ to
1 TeV. Not only inclusive jet rates are affected, but also more exclusive
distributions, both global (like the event shape variables) 
and individual (like energy
and angles) ones. Effects range from a few percent to several tens
of percent, depending on the energy and the observable being studied,
and we have shown cases where such higher-order contributions would impinge
on the experimental measurements of jet quantities.

A careful analysis of actual three-jet data is in order then, which
should not only involve two-loop QCD contributions but also one-loop
EW ones. One {\it caveat} should be borne in mind though concerning the
latter. That is, as emphasised in Ref.~\cite{hadronic}, 
particular care should be devoted to the
treatment of real $W^\pm$ and $Z$ production and decay in the
definition of the jet data sample, as this will
determine whether tree-level $W^\pm$ and $Z$ bremsstrahlung
effects (neglected here) have to be included in the theoretical predictions
through ${\cal O}(\alpha_{\rm S}\alpha_{\rm{EM}}^3)$, which might 
counterbalance in part the 
effects due to one-loop $W^\pm$ and $Z$ virtual exchange. However,
given the cleanliness of leptonic jet data samples, as compared
to hadronic ones, we believe that the former contribution can always
be effectively disentangled in data.

We are now in the process of computing the ${\cal O}(\alpha_{\rm S}\alpha_{\rm{EM}}^3)$
one-loop QCD corrections to $e^+e^-\to q\bar q\gamma$, so that -- upon combining these 
with the computations performed in this paper --  we will eventually be in a position of 
rendering our complete predictions through that order 
of immediate experimental relevance. 

\acknowledgments
SM thanks FP for financial support during a visit to Pavia while
FP thanks SM for the same reason on the occasion of several visits to
Southampton. CMCC acknowledges partial
 financial support from the British Council
in the form of a `Researcher Exchange Programme' award
and from the Royal Society in the form of a `Short Visit to the UK' grant. 
CMCC and FP thank the Galileo Galilei Institute for kind hospitality
and G.~Passarino for useful discussion during the workshop ``Advancing
Collider Physics: from Twistors to Monte Carlos''.

\section*{Appendix}
\setcounter{equation}{0}
\renewcommand{\theequation}{A.\arabic{equation}}
In this Appendix we display the analytic expressions for the
 IR divergent part of the relevant scalar
 integrals used
 in this work. Such divergences occur either in the case in which
all the particles inside the loop are taken to be massless 
(zero-internal-mass integrals), such as box or pentagon graphs in which both
 the exchanged gauge bosons are photons, or graphs which contain one massive
internal particle (one-internal-mass integrals)
but are nevertheless IR divergent, such as box
or pentagon graphs with one photon and one $Z$ boson exchange.
We perform the integrals in $(4-2\epsilon)$ dimensions with subtraction
scale $\mu$.

The zero-internal-mass integrals have been discussed in detail in, e.g.,
Ref.~\cite{bdk}, but for completeness we reproduce the results here, 
introducing some compact notation for integrals that will be used
to describe the one-internal-mass integrals:

\begin{eqnarray}
I_3^{(1,0)}(s) & \equiv & C_0(s,0,0,0,0,0) \ \equiv 
 \frac{1}{\pi^2} \left. \int \frac{d^{(4-2\epsilon)}k}
{k^2 (k+p_1)^2 (k+p_1+p_2)^2}\right|_{p_1^2=p_2^2=0, \ (p_1+p_2)^2=s}
 \nonumber \\ & = &
 \frac{r(\epsilon,\mu)}{\epsilon^2 s } \left(\frac{-s}{\mu^2}\right)^{-\epsilon },
\end{eqnarray}
where
\begin{equation}   r(\epsilon,\mu) \ = \ 
\frac{\Gamma^2(1-\epsilon)\Gamma(1+\epsilon)}{\Gamma(1-2\epsilon)}
\left( \pi \mu^2 \right)^{-\epsilon} 
 \ = \ \Gamma(1+\epsilon)\left( \pi \mu^2 \right)^{-\epsilon} 
 \left(1-\epsilon^2\frac{\pi^2}{6} + \cdots \right).
\end{equation}
\begin{eqnarray}
I_3^{(2,0)}(s,t) & \equiv & C_0(s,t,0,0,0,0) \ \equiv 
  \frac{1}{\pi^2} \left. \int \frac{d^{(4-2\epsilon)}k}
{k^2 (k+p_1)^2 (k+p_1+p_2)^2}\right|_{p_1^2=t, \ p_2^2=0, \ (p_1+p_2)^2=s}
 \nonumber \\ & = &
 \frac{r(\epsilon,\mu)}{\epsilon^2 (s-t)} \left( \left(\frac{-s}{\mu^2}\right)^{-\epsilon }-
\left(\frac{-t}{\mu^2}\right)^{-\epsilon }\right),
\end{eqnarray}

\begin{eqnarray}
I_4^{(1,0)}(s,t,p^2) & \equiv & D_0(s,t,p^2,0,0,0,0,0,0,0) \ \equiv 
  \frac{1}{\pi^2} \int \frac{d^{(4-2\epsilon)}k}
{k^2 (k+p_1)^2 (k+p_1+p_2)^2  (k-p_4)^2}
 \nonumber \\ & = &
 \frac{r(\epsilon,\mu)}{st} \left\{ \frac{2}{\epsilon^2} \left(
\left(\frac{-s}{\mu^2}\right)^{-\epsilon}+
\left(\frac{-t}{\mu^2}\right)^{-\epsilon}-
\left(\frac{-p^2}{\mu^2}\right)^{-\epsilon}
\right) \right. \nonumber \\ & & \left.
 - 2 \mathrm{Li}_2\left(1-\frac{p^2}{s}\right)
 - 2 \mathrm{Li}_2\left(1-\frac{p^2}{t}\right)
 - \ln \left( \frac{t}{s}\right)^2 - \frac{\pi^2}{3}
 \right\}.
\end{eqnarray}
For the above box integral, 
$p_1^2=p^2, \ p_i^2=0 \, (i>1), \ (p_1+p_2)^2=s, \  (p_2+p_3)^2=t$.

\begin{eqnarray}
I_5^{(0,0)}(s_{12},s_{23},s_{34},s_{45},s_{51}) & \equiv &
E_0(s_{12},s_{23},s_{34},s_{45},s_{51},0,0,0,0,0,0,0,0,0,0) \ \equiv 
\nonumber \\  
   & & \hspace*{-3cm}
 \frac{1}{\pi^2}  \left. \int \frac{d^{(4-2\epsilon)}k}
{k^2 (k+p_1)^2 (k+p_1+p_2)^2  (k-p_4-p_5)^2 (k-p_5)^2} 
\right|_{p_i^2=0, \ s_{ij}=2p_i\cdot p_j}
 \nonumber =  \\ &  & \hspace*{-3cm}
\frac{-1}{s_{23}s_{34}s_{45}}
 \left\{ 
\frac{r(\epsilon,\mu)}{\epsilon^2} 
\left(\frac{-s_{23}s_{34}s_{45}}{s_{51}s_{12} \mu^2} \right)^{-\epsilon}
+2 \mathrm{Li}_2(\left(1-\frac{s_{23}}{s_{51}} \right)
+2 \mathrm{Li}_2(\left(1-\frac{s_{45}}{s_{12}}\right)-\frac{\pi^2}{6}
 \right\}  \nonumber   \\ &  & 
\mbox{ + cyclic permutations.} 
\end{eqnarray}

For the one-internal-mass-integrals, we begin with the IR
divergent triangle integrals, i.e., those in which the internal massive
propagator is adjacent to (as opposed to opposite) an external ``mass''.

\begin{eqnarray}
I_3^{(1,1)}(s,M^2) & \equiv & C_0(s,0,0,M^2,0,0)  \nonumber \\ &  \equiv & 
 \frac{1}{\pi^2} \left. \int \frac{d^{(4-2\epsilon)}k}
{(k^2-M^2) (k+p_1)^2 (k+p_1+p_2)^2}\right|_{p_1^2=p_2^2=0, \ (p_1+p_2)^2=s}
 \nonumber \\ & = &
 \frac{r(\epsilon,\mu)}{\epsilon^2 s } \left(
\left( \frac{(M^2-s)}{\mu^2}\right)^{-\epsilon }
- \left( \frac{M^2}{\mu^2}\right)^{-\epsilon } \right) - \frac{1}{s}
 \mathrm{Li}_2\left( \frac{s}{(s-M^2)}    \right).
\end{eqnarray}

\begin{eqnarray}
I_3^{(2,1)}(s,t,M^2) & \equiv & C_0(s,t,0,0,0,0) \nonumber \\ & \equiv & 
 \frac{1}{\pi^2} \left. \int \frac{d^{(4-2\epsilon)}k}
{(k^2-M^2)
 (k+p_1)^2 (k+p_1+p_2)^2}\right|_{p_1^2=t, \ p_2^2=0, \ (p_1+p_2)^2=s}
  \nonumber \\   & & \hspace*{-3cm} \ = \ 
 \frac{r(\epsilon,\mu)}{ (s-t)} \left\{ \frac{1}{\epsilon^2} \left(
 \left(\frac{M^2-s}{\mu^2}\right)^{-\epsilon }-
\left(\frac{M^2-t}{\mu^2}\right)^{-\epsilon } \right)
 -\mathrm{Li}_2\left( \frac{s}{(s-M^2)}    \right)
 +\mathrm{Li}_2\left( \frac{t}{(t-M^2)}    \right)
\right\}. \nonumber \\ & &
\end{eqnarray}

We now show how the IR divergent part of
an $n$-point integral can be obtained in terms of 
$(n-1)$-point integrals. This method is similar to that of
Ref.~\cite{ditt_ir}, but the resulting expressions are somewhat simpler. 
Consider an $n$-point integral with general incoming external momenta
$p_i \ (i=1 \cdots n)$ and a massive propagator on the line $j_0$,
\begin{equation}
I_n(p_1 \cdots p_n, \cdots m_{j_0} \cdots) \ \equiv \ 
 \frac{1}{\pi^2}  \int \frac{d^{(4-2\epsilon)}k}
{\prod_{j=0}^n \{ (k+q_j)^2 -m_j^2 \} }, \end{equation}
where
\begin{equation} q_j \ = \ \sum_{i=1}^j p_i, \ \ q_0=0, \end{equation} 
and at least $m_{j_0}$ is non-zero  (there may, in general, be other
internal finite masses, $m_j$).

After Feynman parameterisation the integral over the loop-momentum
$k$ may be performed yielding
\begin{equation}
I_n(p_1 \cdots p_n, \cdots m_{j_0} \cdots) \ = \ 
\Gamma(n+\epsilon-2)
\int \frac{ \prod_j d \alpha_j \, \delta(1-\sum_j \alpha_j)}
{\left[ 
\sum_{i,j}(q_i-q_j)^2 \alpha_i \alpha_j - m_{j_0}^2 \alpha_{j_0}
-\cdots \right]^{n-2+\epsilon}}.
\end{equation}
The integration over the Feynman parameters can only give rise to
an IR divergence when $\alpha_{j_0}=0$. In other words, the 
integral
\begin{equation}\int \frac{ \prod_j d \alpha_j \, \delta(1-\sum_j \alpha_j)
\alpha_{j_0}}
{\left[ 
\sum_{i,j}(q_i-q_j)^2 \alpha_i \alpha_j - m_{j_0}^2 \alpha_{j_0}
-\cdots \right]^{n-2+\epsilon}}\end{equation}
is finite.

The term $\alpha_{j_0}$ in the numerator can be generated
by considering the integral
\begin{equation}
 \frac{1}{\pi^2}  \int \frac{d^{(4-2\epsilon)}k \, 2 G^{-1}_{j_0l}
  q_l \cdot k}
{\prod_{j=0}^n \{ (k+q_j)^2 -m_j^2 \} }, \label{llama}
\end{equation}
 where the Gram matrix, $G$, is defined by
\begin{equation} G_{ij} \ = \ 2 q_i \cdot q_j. \end{equation}
Now, by using the ``pinch'' relations
\begin{equation} 2 q_l \cdot k \ = \ \left( (k+q_l)^2-m_l^2 \right)
   - (k^2-m_0^2) + m_l^2-m_0^2-q_l^2 \end{equation}
and the fact that the integral (\ref{llama}) is IR finite,
we obtain the result for the IR divergence part of the integral
$I_n$,
\begin{equation}
\left\{ \sum_l G^{-1}_{j_0l} \left(m_l^2-m_o^2-q_l^2 \right) \right\}
I_n(p_1 \cdots p_n, \cdots m_{j_0} \cdots)|_{\rm{IR}} \ = \
  \sum_l G^{-1}_{j_0l} \left(I_{(n-1)}^{\{l\}}-
I_{(n-1)}^{\{0\}} \right) \label{alpaca}, \end{equation}
where $I_{(n-1)}^{\{l\}}$ is the  $(n-1)$-point
 integral obtained by  ``pinching out'' the $l$-${\mathrm{th}}$
propagator from the $n$-point integral, $I_n$.

Some of the integrals on the RHS of Eq.~(\ref{alpaca})
may be IR convergent and may therefore be dropped.
This process may then be iterated so that eventually
the finite part of any integral with one internal mass can
be expressed in terms of integrals with zero internal masses
and the IR divergent triangle integrals.

The IR convergent contributions can then be obtained numerically,
by taking the difference between the original  
IR divergent integrals on the  RHS of Eq.~(\ref{alpaca})
and regularising the divergences by inserting a small mass for all
the internal propagators and then checking that the result is 
insensitive to the value taken for this regulator mass, provided it
is much smaller than any of the other masses or momenta but not
so small that it introduces numerical instabilities.

For the integrals required in this calculation, we have two
IR  box integrals with one non-zero internal mass and one non-zero
external mass, depending on whether the external mass is opposite (a)
or adjacent (b) to the massive propagator. 

\begin{eqnarray}
I_4^{(a)}(s,t,p^2,M^2) &\equiv &
D_0(s,t,p^2,0,0,0,0,0,M^2,0) \nonumber \\ & \equiv & 
 \frac{1}{\pi^2}  \int \frac{d^{(4-2\epsilon)}k}
{k^2 (k+p_1)^2 ((k+p_1+p_2)^2-M^2)  (k-p_4)^2},
\end{eqnarray}
with $p_1^2=p^2, \ p_i^2=0 \, (i>1), \ (p_1+p_2)^2=s, \  (p_2+p_3)^2=t$.
\begin{eqnarray}
I_4^{(b)}(s,t,p^2,M^2) &\equiv &
D_0(s,t,0,p^2,0,0,0,0,M^2,0) \nonumber \\ & \equiv & 
 \frac{1}{\pi^2}  \int \frac{d^{(4-2\epsilon)}k}
{k^2 (k+p_1)^2 ((k+p_1+p_2)^2-M^2)  (k-p_4)^2},
\end{eqnarray}
with $p_1^2=0, \ p_2^2=p^2 \ p_i^2=0 \, (i>2), 
\ (p_1+p_2)^2=s, \  (p_2+p_3)^2=t$.

Using the above technique
 the IR divergent parts of these integrals are given by
\begin{equation}
I_4^{(a)}(s,t,p^2,M^2)_{\rm{IR}} \ = \ 
\frac{s I_3^{(1,1)}(s,M^2)+(t-p^2)I_3^{(2,0)}(t,p^2)}
{st-M^2(t-p^2)} \end{equation}
and
\begin{equation}
I_4^{(b)}(s,t,p^2,M^2)_{\rm{IR}} \ = \ 
\frac{s I_3^{(1,1)}(s,M^2)+tI_3^{(1,0)}(t)+(s-p^2)I_3^{(2,1)}(s,p^2,M^2)}
{t(s-M^2)}. \end{equation}

Finally we have the pentagon integral with no external masses and one 
internal mass, chosen arbitrarily to be the line between incoming
momenta $p_2$ and $p_3$,
\begin{eqnarray}
I_5^{(0,1)}(s_{12},s_{23},s_{34},s_{45},s_{51},M^2) & \equiv &
E_0(s_{12},s_{23},s_{34},s_{45},s_{51},0,0,0,0,0,0,0,M^2,0,0) \ \equiv 
\nonumber \\  
   & & \hspace*{-5.5cm}
 \frac{1}{\pi^2}  \left. \int \frac{d^{(4-2\epsilon)}k}
{k^2 (k+p_1)^2 ((k+p_1+p_2)^2-M^2)  k-p_4-p_5)^2 (k-p_5)^2} 
\right|_{p_i^2=0, \ s_{ij}=2p_i\cdot p_j}. \end{eqnarray}

The IR divergent part of this integral is found to be
\begin{eqnarray}
I_5^{0,1}(s_{12},s_{23},s_{34},s_{45},s_{51},M^2)_{\rm{IR}} & = &
 \frac{1}{s_{51}s_{45} (s_{14}s_{25}+s_{12}s_{45}+2s_{12}s_{14}-s_{51}s_{24}
   -2 M^2 s_{14})} \nonumber \\
 &  & \hspace*{-6.5cm}   \ \times \ 
\left\{ \left(s_{51}^2s_{24} - s_{14}s_{51}s_{25} + s_{14}s_{51}s_{24} - 
       s_{14}^2s_{25} - s_{12}s_{51}s_{45} + s_{12}s_{14}s_{45} \right)
 I_4^{(a)}(s_{12},s_{23},s_{45},M^2) \right. \nonumber \\
 &  & \hspace*{-5.8cm} \ + \ 
\left. \left( s_{12}s_{51}s_{45}
 - s_{51}^2s_{24} + s_{14}s_{51}s_{25} 
 \right)
 I_4^{(b)}(s_{12},s_{51},s_{34},M^2) \right.
+2 s_{14}s_{51}s_{45} I_4^{(1,0)}(s_{51},s_{45},s_{23}) \nonumber \\
  &  & \hspace*{-5.8cm} \ + \
\left. \left(
s_{51}s_{24}s_{45} + s_{14}s_{25}s_{45} - s_{14}s_{51}s_{24} 
+ s_{14}^2s_{25}
          - s_{12}s_{45}^2 - s_{12}s_{14}s_{45}
 \right)
 I_4^{(a)}(s_{34},s_{23},s_{51},M^2) \right. \nonumber \\
 &  & \hspace*{-5.8cm} \ + \
\left. \left(s_{12} s_{45}^2
- s_{51}s_{24}s_{45} - s_{14}s_{25}s_{45} - 2s_{14}s_{51}s_{45}  
 \right)
 I_4^{(b)}(s_{34},s_{45},s_{12},M^2) \right\}.
  \end{eqnarray}


\newpage
\begin{center}
{\LARGE\bf Erratum to:\\
\vskip 0.1cm
Full One-loop Electro-Weak Corrections to Three-jet Observables at the 
$Z$ Pole and Beyond\\
\vskip 0.1cm
[JHEP 0903:047,2009]}
\end{center}
\vskip 0.6cm
{\bf C.M. Carloni-Calame$^{1,2}$, S. Moretti$^{2}$, F. Piccinini$^{3}$ and D.A. Ross$^{2}$}\\
\vskip 0.1cm
\noindent
{\it {$^1$ INFN, via E. Fermi 40, Frascati, Italy}
\\
{$^2$ School of Physics and Astronomy, University of Southampton}
{Highfield, Southampton SO17 1BJ, UK}
\\
{$^3$ INFN - Sezione di Pavia, Via Bassi 6, 27100 Pavia, Italy}}
\vskip 1cm
\noindent
All the integrated and differential cross-sections presented in the
original paper have been mistakenly multiplied by an overall factor of
$1/3$. Furthermore, besides the cuts described in section
\ref{Sec:Results}, an additional cut was
imposed to obtain the results which is not correctly stated in the
paper: namely, the energy of each jet is
required to be larger than $\mbox{min}(5\mbox{ GeV},0.05\times\sqrt{s})$.

However, the relative effects due to the impact of the full one-loop EW  
terms of ${\cal O}(\alpha_{\rm S}\alpha_{\rm{EM}}^3)$ remain unchanged
as well as the conclusions of our study.

\vskip 0.5cm
\noindent
{\Large\bf Acknowledgments}
\vskip 0.2cm
\noindent
We gratefully acknowledge an email exchange with Ansgar
Denner, Stefan Dittmaier, Thomas Gehrmann and
Christian Kurz [1] and thank them for pointing out the
discrepancy with their independent
calculation.
\vskip 0.5cm
\noindent
{\Large\bf References}
\vskip 0.2cm
\noindent
[1] Thomas Gehrmann, Christian Kurz, Ansgar Denner, Stefan Dittmaier,
private communication. Paper in preparation.


\begin{thebibliography}{99}
%
\bibitem{Kuroda:1991wn}
M.~Kuroda, G.~Moultaka and D.~Schildknecht,
\npb{350} {1991} {25};\\
G.~Degrassi and A.~Sirlin,
\prd{46} {1992} {3104};\\
A.~Denner, S.~Dittmaier and R.~Schuster,
\npb{452} {1995} {80};\\
A.~Denner, S.~Dittmaier and T.~Hahn,
\prd{56} {1997} {117};\\
A.~Denner and T.~Hahn,
\npb{525} {1998} {27}.

\bibitem{Beenakker:1993tt}
W.~Beenakker, A.~Denner, S.~Dittmaier, R.~Mertig and T.~Sack,
\npb{410} {1993} {245};
~\plb{317} {1993} {622}.

\bibitem{Ciafaloni:1999xg}
P.~Ciafaloni and D.~Comelli,
\plb{446} {1999} {278}.

\bibitem{Denner:2000jv}
   A.~Denner and S.~Pozzorini,
   Eur.\ Phys.\ J.\  C {\bf 18} (2001) 461.

\bibitem{KLN} T. Kinoshita, {J. Math. Phys.\ }{\bf 3} (1962) 650;\\
T.D. Lee and M. Nauenberg, {Phys.\ Rev.\ }{\bf 133} (1964) 1549.

\bibitem{BN} F. Bloch and A. Nordsieck, 
{Phys.\ Rev.\ }{\bf 52} (1937) 54.

\bibitem{Ciafaloni:2000df}
M.~Ciafaloni, P.~Ciafaloni and D.~Comelli,
\prl{84} {2000} {4810}.


\bibitem{2jet} See, e.g.:
M. Consoli and W. Hollik (conveners), in proceedings of the workshop
`$Z$ Physics at LEP1' (G. Altarelli, R. Kleiss and C.
Verzegnassi, eds.), preprint CERN-89-08, 21 September 1989
(and references therein).

\bibitem{4jet} V.A. Khoze, D.J. Miller, S. Moretti and 
W.J. Stirling, \jhep {07} {1999} {014}.

\bibitem{oldpapers}
  E.~Maina, S.~Moretti, M.R.~Nolten and D.A.~Ross,
  arXiv:hep-ph/0407150;
  arXiv:hep-ph/0403269;
  JHEP {\bf 0304} (2003) 056.


\bibitem{Baur-rad}
  U.~Baur,
  Phys.\ Rev.\  {\bf D75} (2007) 013005.


\bibitem{LCs}
A.~Djouadi, J.~Lykken, K.~Monig, Y.~Okada, M.J.~Oreglia and S.~Yamashita,
  arXiv:0709.1893 [hep-ph];\\
K.~Abe {\it et al.}, [The ACFA Linear Collider Working Group],
{arXiv:hep-ph/0109166};\\
T.~Abe {\it et al.}, [The American Linear Collider Working Group],
{arXiv:hep-ex/0106055}; {arXiv:hep-ex/0106056}; {arXiv:hep-ex/0106057};
{arXiv:hep-ex/0106058};\\
J.A. Aguilar-Saavedra {\it et al.}, [The 
ECFA/DESY LC Physics Working Group],  {arXiv:hep-ph/0106315};\\
G. Guignard (editor), [The CLIC Study Team], preprint CERN-2000-008 (2000).

\bibitem{Dissertori} 
G. Dissertori, talk presented at the `XXXI  International Conference
on High Energy Physics', Amsterdam, 24-31 July 2002, preprint  
{arXiv:hep-ex/0209070} (and reference therein).
 
\bibitem{Winter} M. Winter, LC Note LC-PHSM-2001-016, February 2001
(and references therein).

\bibitem{QCD2Loops}
%
  A.~Gehrmann-De Ridder, T.~Gehrmann, E.W.N.~Glover and G.~Heinrich,
  arXiv:0802.0813 [hep-ph];
%
  arXiv:0801.2680 [hep-ph];
%
  JHEP {\bf 0712} (2007) 094;
%
  JHEP {\bf 0711} (2007) 058;
%
  arXiv:0709.4221 [hep-ph];
%
  arXiv:0709.1608 [hep-ph];
%
  0707.1285 [hep-ph];
  Nucl.\ Phys.\ Proc.\ Suppl.\  {\bf 160} (2006)  190;\\
  A.~Gehrmann-De Ridder, T.~Gehrmann and E.W.N.~Glover,
  Nucl.\ Phys.\ Proc.\ Suppl.\  {\bf 135} (2004)  97;\\
L.W. Garland, T. Gehrmann, E.W.N. Glover, A. Koukoutsakis and E. Remiddi, 
{Nucl. Phys.} {\bf B642} (2002) 227;~
 {Nucl. Phys.} {\bf B627} (2002) 107.

\bibitem{ERT} R.K. Ellis, D.A. Ross and A.E. Terrano, \npb{178} {1981} {421}.

\bibitem{3j1a}
  S.~Moretti, R.~Munoz-Tapia and J.B.~Tausk,
  arXiv:hep-ph/9609206;\\
  S.~Moretti and J.B.~Tausk,
  Z.\ Phys.\   {\bf C69} (1996)  635.

\bibitem{RADCOR} C.M. Carloni-Calame, S. Moretti, F. Piccinini and D.A. Ross,
talk presented at ``RADCOR 2007:
8th International Symposium on Radiative Corrections'',
Florence, Italy, October 1-5, 2007, arXiv:0804.1657 [hep-ph]. 

\bibitem{Maina:2002wz}
E.~Maina, S.~Moretti and D.A.~Ross,
last paper in \cite{oldpapers}.


\bibitem{PV} G. Passarino and M.J.G. Veltman, {Nucl. Phys.}
{\bf B160} (1979) 151. 

\bibitem{FORM} J.A.M. Vermaseren, math-ph/0010025.

\bibitem{complexmass} A.~Denner, S.~Dittmaier, M.~Roth and D.~Wackeroth, 
\npb{560} {1999} {33}; \\ 
A.~Denner, S.~Dittmaier, M.~Roth and L.H.~Wieders, 
\npb{724} {2005} {247}; \\ 
A.~Denner and S.~Dittmaier, Nucl. Phys. Proc. Suppl. 
{\bf 160} (2006) 22.


\bibitem{pivovarov} A.A. Pivovarov, {Phys. Atom. Nucl} {\bf 65} (2000) 1319.
 
\bibitem{VG} 
M. Green and M.J.G. Veltman, Nucl. Phys. {\bf B169} (1980) 137 
[{Erratum}, {\it ibidem} {\bf B175} (1980) 547].

\bibitem{looptools} T. Hahn and M. Perez-Victoria, {Comput. Phys. Commun.} {\bf 118} 
  (1999) 153.

\bibitem{DD} A.~Denner and S.~Dittmaier, {Nucl. Phys.} {\bf B658} (2003) 175.

\bibitem{FF1.9} G.J. van Oldenborgh, \cpc {66} {1991} {1}. 

\bibitem{ALPHA} F. Caravaglios and M. Moretti, {Phys. Lett.} {\bf B538} (1995) 332.

\bibitem{SL} T. Stelzer and W.F. Long, {Nucl. Phys. Proc. Suppl. } {\bf 37B} 
(1994) 158.

\bibitem{BMM}
A.~Ballestrero, E.~Maina and S.~Moretti,
{Phys.\ Lett.}  {\bf B294} (1992) 425;
{Nucl. Phys.} {\bf B415} (1994) 265.


\bibitem{bbatnlo} G.~Balossini, 
C.M.~Carloni Calame, G.~Montagna, O.~Nicrosini and F.~Piccinini, {Nucl. Phys.} {\bf B758} (2006) 227.

\bibitem{sf} E.A.~Kuraev and V.S.~Fadin, Sov. J. Nucl. Phys. {\bf 41} (1985)
466; \\
G.~Altarelli and G.~Martinelli, in {\it Physics at LEP} (J.~Ellis and
R.~Peccei, eds.), CERN Report {86-02} (Geneva, 1986), vol.~1, p.~47; \\
O.~Nicrosini and L.~Trentadue, Phys. Lett. {\bf B196} (1987) 551; Z. Phys. {\bf
C39} (1988) 479.

\bibitem{MNP} G. Montagna, O. Nicrosini and F. Piccinini, \plb{385} {1996} {348}. 
\bibitem{tHV} G.~'t Hooft and M.J.G.~Veltman, \npb{153} {1979} {365}.

\bibitem{fixedwidth} See for instance W.~Beenakker et al., \npb {500} {1997} {255}.

\bibitem{cambridge} Yu.L.~Dokshitzer, G.D.~Leder, S.~Moretti and
  B.R.~Webber, \jhep {08} {1997} {001};\\
S. Moretti, L. L\"onnblad and T. Sj\"ostrand,
\jhep {08} {1998} {001}.

\bibitem{thrust} S.~Brandt, Ch.~Peyrou, R.~Sosnowski and A.~Wroblewski, 
\plb {12} {1964} {57};\\
E.~Farhi, \prl {39} {1977} {1587}.

\bibitem{spherocity} H.~Georgi and M.~Machacek, \prl {39} {1977} {1237}.

\bibitem{cparam} R.K.~Ellis, D.A.~Ross and A.E.~Terrano, 
\npb {178} {1981} {421}.

\bibitem{oblateness} D.P.~Barber et al., (MARK-J Collaboration), 
\prl {43} {1979} {830}.

\bibitem{ALEPH} ALEPH Collaboration, 
`QCD studies with $e^+e^-$ annihilation data at 189 GeV',
ALEPH 99-023, CONF 99-018, June 1999 (pPrepared for 
the International Europhysics Conference on High-Energy Physics (EPS-HEP 99), Tampere, Finland, 15-21 Jul 1999).

\bibitem{kunsztnason} See, e.g.: Z. Kunszt and P. Nason, (conveners), 
in proceedings of the workshop
`$Z$ Physics at LEP1' (G. Altarelli, R. Kleiss and C.
Verzegnassi, eds.), preprint CERN-89-08, 21 September 1989 
(and references therein).

\bibitem{hadronic}
S. Moretti, M.R. Nolten and D.A. Ross, 
Phys. Rev. {\bf D74} (2006) 097301; 
Phys. Lett. {\bf B643} (2006) 86; 
Phys. Lett. {\bf B639} (2006) 513 
[Erratum, {\it ibidem} {\bf B660} (2008) 607]; 
Nucl. Phys. {\bf B759} (2006) 50;\\
E. Maina, S. Moretti, M.R. Nolten and D.A. Ross, 
Phys. Lett. {\bf B570} (2003) 205. 


\bibitem{bdk}
Z.~Bern, L.J.~Dixon and D.A. Kosower, 
{Nucl. Phys.} {\bf B412} (1994) 751.

\bibitem{ditt_ir}
S.~Dittmaier,
{Nucl. Phys.} {\bf B675} (2003) 447.

\end{thebibliography}
\end{document}